\documentstyle[12pt,aps,pra]{revtex}
\begin{document}
\title{Transport Coefficients of the Anderson Model via the
Numerical Renormalization Group}
\author
{T. A. Costi\cite{tac-present,tac-permanent}, A.C. Hewson}
\address
{Department of Mathematics, Imperial College, London}
\author
{V. Zlati\'{c}\cite{vz-present}}
\address
{Institute of Physics, University of Zagreb, Croatia}
\date{\today}
\maketitle

\begin{abstract}
The transport coefficients of the Anderson model are calculated by
extending Wilson's numerical renormalization group method to finite
temperature Green's functions. Accurate results for the frequency and
temperature dependence of the single--particle spectral densities and
transport time $\tau(\omega,T)$ are obtained and used to extract the
temperature dependence of the transport coefficients in the strong
correlation limit of the Anderson model.  Results are obtained for
values of the local level position ranging from the Kondo regime to
the mixed valent and empty orbital regimes. The low temperature
anomalies in the resistivity, $\rho(T)$, thermopower, $S(T)$, thermal
conductivity $\kappa(T)$ and Hall coefficient, $R_{H}(T)$, are
discussed in terms of the behaviour of the spectral densities. All
quantities exhibit the expected Fermi liquid behaviour at low
temperature, $\rho(T)=\rho_{0}(1-c (T/T_{K})^2)$, $S(T)\sim \gamma T$,
$\kappa(T)/\alpha\;T = 1 +\beta (T/T_{K})^{2}$, $R_{H}(T) =
-R_{\infty}(1-\delta (T/T_{K})^{2})$. Analytic results based on Fermi
liquid theory are derived here for the first time for $\beta$ and the
numerical results are shown to be consistent with this coefficient.
The range of temperatures over which universal behaviour extends is
also discussed.  Scattering of conduction electrons in higher, $l>0$,
angular momentum channels is also considered and an expression is
derived for the corresponding transport time and used to discuss the
influence of the interference terms between the resonant $l=0$ and
non--resonant $l=1$ channels on the transport properties.  The
presence of non--resonant scattering is shown to be particularly
important for the thermopower at half--filling, where the sign of the
thermopower can depend sensitively on the non--resonant phase shift.
Finally the relation of the results to experiment is discussed.
\end{abstract}

\section*{Introduction}
In this paper we present accurate results for the transport
coefficients of the Anderson model obtained by extending the numerical
renormalization group method
\cite{wilson.75,krishnamurthy.80} to
 the calculation of finite temperature Green's functions.  The
Anderson model has been used extensively to interpret the properties
of dilute magnetic alloys and a number of the local properties of
heavy fermion compounds. It is also of use in discussing the
properties of concentrated Kondo systems in cases where alloying or
disorder inhibit the coherence effects between the magnetic ions.
Although many of the properties of this model are now well understood
\cite{hewson.93}, the temperature dependence of the transport coefficients has
proved to be particularly difficult to calculate reliably. The
transport coefficients require accurate expressions for both the
temperature and frequency dependence of the impurity Green's function,
a quantity which is difficult to calculate in the strong correlation
limit of the Anderson model. Here we calculate this quantity from
Wilson's numerical renormalization group method
\cite{wilson.75} which is non--perturbative and is therefore accurate for
arbitrarily large Coulomb interactions. This method has played a
crucial role in forming our current understanding of the Anderson
model. It was first applied to the Kondo model by Wilson
\cite{wilson.75} and subsequently to the Anderson model by
Krishnamurthy et al \cite{krishnamurthy.80}. The two models are
related via the Schrieffer--Wolff transformation \cite{schrieffer.66},
with the Kondo model describing the low--energy physics of the
Anderson model in the strong correlation Kondo regime.  The
application of the numerical renormalization group to these models
yielded the elementary excitations, thermodynamics, fixed points and
effective Hamiltonians around the fixed points
\cite{wilson.75,krishnamurthy.80}.  Dynamic and transport properties were not
calculated. The calculations showed that the Kondo model has two fixed
points which characterize its physical properties. The local moment
fixed point which describes the high temperature regime and in which
the conduction electrons couple weakly to the impurity moment and a
strong coupling fixed point which describes the low temperature regime
in which the impurity moment is quenched and the excitations are those
of a local Fermi liquid. The parameter space of the Anderson model is
larger and the calculations showed that in addition to the local
moment and strong coupling fixed points there are two additional fixed
points. The most important of these is the valence fluctuation fixed
point and is characteristic of the asymmetric model. In the valence
fluctuation regime, charge fluctuations become important and the
properties correspond to a model with a strongly renormalized
temperature dependent resonant level
\cite{krishnamurthy.80}. In this paper we describe the transport coefficients
and their relation to the corresponding spectral densities in these
different regimes, pointing out the characteristic features which
arise in each case.

In contrast to thermodynamic properties, which have been obtained
exactly by the Bethe Ansatz \cite{andrei.83,tsvelick.83} and numerical
renormalization group
\cite{wilson.75,krishnamurthy.80}, the calculation of transport and dynamic
properties have relied on approximate methods. Finite order
perturbation in $U$ calculations give accurate results in the Fermi
liquid regime for spectral densities and thermodynamic properties up
to $U/\pi\Delta\approx 2.5$, where $U$ is the local Coulomb repulsion
and $\Delta$ is the resonant level width in the Anderson model
\cite{horvatic.87,yamada.76}.  However, in the local moment regime
$U>>\Delta$ for $T>>T_{K}$ where properties depend on logarithmic
terms this approach breaks down.  Quantum Monte Carlo approaches
\cite{silver.89,silver.90} become increasingly less accurate for larger values
of $U$ and lower temperatures. So far transport coefficients via this
method have been calculated for just the symmetric Anderson model and
for $U/\pi\Delta
\le 3$ \cite{jarrell.91}. Transport coefficients have also been obtained for
larger degeneracies of the local level $N$ via the non--crossing
approximation
\cite{bickers.87,qin.91}. A problem with this method is that it fails to
satisfy
the Fermi liquid relations at zero temperature \cite{mh.84}. In a
finite magnetic field the transport coefficients for the Kondo model
were discussed in
\cite{keiter.76} on the basis of the Nagaoka integral equations.  Recently a
comprehensive and highly accurate approach to the calculation of
dynamic properties of magnetic impurity models has been developed by
extending the numerical renormalization group approach
\cite{frota.86,sakai.89,sakai.90,costi.90,costi.91,costi.92a,costi.92b}.  This
overcomes the above mentioned difficulties with the approximate
methods.  Accurate results in all regimes have been obtained for
single--particle spectral densities at both zero
\cite{sakai.89,sakai.90,costi.90,costi.91,costi.92a} and finite
temperature \cite{costi.92b,costi.93a}. These satisfy all the sum
rules and Fermi liquid relations.  In the next section we introduce
the Anderson model including terms which model the scattering of
conduction electrons in higher ($l>0$) angular momentum channels.  The
transport coefficients are defined in terms of the transport time for
conduction electrons scattering in both the the resonant and
non--resonant channels (the transport time incorporating non--resonant
scattering of conduction electrons is derived in Appendix~1). The
numerical renormalization group and its use in extracting finite
temperature Greens's functions and spectral densities is then
described. We also give analytic calculations for the low temperature
behaviour of the transport coefficients based on Fermi liquid theory.
These are used as a check on the accuracy of the numerical results.
Finally we present the conclusions and indicate the relevance of the
results to experiment.

\section*{Model, Transport Properties and Method}

\subsection*{The model}

The Anderson model including non--resonant scattering of conduction
electrons in $l> 0$ channels is given by the following Hamiltonian
\begin{eqnarray}
H & = & H_{imp} + H_{hyb} + H_{c},\label{eq:model}\\ H_{imp} & = &
\sum_{\sigma}\epsilon_{0}c_{0\sigma}^{\dagger}c_{0\sigma} +
Un_{0{\uparrow}}n_{0{\downarrow}}+
\sum_{l>0}\sum_{m=-l}^{+l}\epsilon_{l}c_{lm}^{\dagger}c_{lm},\nonumber\\
H_{hyb} & = & \sum_{k\sigma}(V_{0k}c_{k\sigma}^{\dagger}c_{0\sigma} +
H.c.) +
\sum_{k\sigma\;l>0}\sum_{m=-l}^{+l}(V_{lmk}c_{k\sigma}^{\dagger}c_{lm} +
H.c.),\nonumber\\ H_{c} & = &
\sum_{k\sigma}\epsilon_{k}c_{k\sigma}^{\dagger}c_{k\sigma}.\nonumber
\end{eqnarray}
The first term $H_{imp}$ represents the impurity and is coupled to the
conduction electrons $H_{c}$ via the hybridization term $H_{hyb}$.
The scattering of conduction electrons in $l>0$ channels is modelled
by including uncorrelated levels, $\epsilon_{l}$, hybridizing with the
conduction electrons.  This is equivalent to taking into account phase
shifts $\eta_{l}$ with $l>0$ for the conduction electrons in addition
to the usual $l=0$ resonant phase shift.  The charge neutrality
condition requires that these phase shifts satisfy the Friedel sum
rule \cite{friedel.52}
\begin{equation}
Z = \sum_{l=0}2(2l+1)\frac{\eta_{l}}{\pi},\label{eq:charge-neutrality}
\end{equation}
where $Z$ is the excess charge on the impurity.

The many body effects arise from the strong Coulomb repulsion between
the electrons in the impurity $l=0$ level.

\subsection*{Transport coefficients}

Assuming that the conduction electrons scatter incoherently from a
small concentration, $n_{i}<<1$, of magnetic impurities, linear
response theory allows the transport coefficients to be expressed in
terms of the transport integrals
\cite{mahan.81}
\begin{equation}
L_{ml}=\int_{-\infty}^{+\infty}{\left (-{\partial f(\omega)
\over \partial \omega}\right )}
\tau^{l}(\omega){(\omega-\mu)}^{m}d\omega,\label{eq:transport-integrals}
\end{equation}
where $\mu$ is the chemical potential, and $\tau(\omega,T)$ is the
transport time. The resistivity, $\rho(T)$, thermoelectric power,
$S(T)$, thermal conductivity $\kappa(T)$ and Hall coefficient
$R_{H}(T)$ are given in terms of these by
\begin{equation}
\rho(T) = {1 \over {e^{2} L_{01}}} = {1 \over {e^{2}}}
{ 1
\over
\int_{}^{}\tau(\omega,T)
\left(-{{\partial f}\over {\partial\omega}}\right)\,
d\omega},
\label{eq:resistivity}
\end{equation}

\begin{equation}
S(T) = -{1 \over {|e| T}}{L_{11} \over L_{01}} = -{1 \over {|e|T}} {
\int_{}^{}\omega\tau(\omega,T)
\left(-{{\partial f}\over {\partial\omega}}\right)\,
d\omega
\over
\int_{}^{}\tau(\omega,T)
\left(-{{\partial f}\over {\partial\omega}}\right)\,
d\omega},
\label{eq:thermo}
\end{equation}
\begin{equation}
\kappa(T) = {1 \over T}\left\{ {L_{21} - {L_{11}^{2} \over
L_{01}}}\right\},\label{eq:thermal-cond}
\end{equation}
\begin{equation}
R_{H}(T) = -R_{\infty}{L_{02} \over {
L_{01}^{2}}},\label{eq:hall-coefficient}
\end{equation}
where $R_{\infty}^{-1}=n_{i}|e|c$.  In the absence of non--resonant
scattering, the transport time, $\tau_{0}$, (see Appendix~1 for
constant factors such as $n_{i}$) is given by
\begin{equation}
\frac{1}{\tau_{0}(\omega,T)}
= \Delta\rho_{0}(\omega, T),\label{eq:tau-resonant-only}
\end{equation}
where $\Delta$ is the resonant level width and $\rho_{0}$ is the
single--particle spectral density. The latter is given in terms of the
resonant level Green's function
$G_{0}(\omega,T)=<<c_{0\sigma};c_{0\sigma}^{\dagger}>>$ and
self--energy $\Sigma(\omega,T) = \Sigma^{R}+i\Sigma^{I}$ by
\begin{equation}
\rho_{0}(\omega,T) = -{{1}\over {\pi}} Im\thinspace G_{0}(\omega,T) =
\frac{1}{\pi}\frac{ (\Delta - \Sigma^{I}(\omega))}{((\omega - \epsilon_{0}-
\Sigma^{R}(\omega))^2 + (\Delta -
\Sigma^{I}(\omega))^{2})}.\label{eq:spec-density}
\end{equation}

 The transport time in the presence of non--resonant scattering is
derived from the Kubo formula for the conductivity in Appendix~1. In
contrast to the case of resonant scattering only (see e.g.
\cite{bickers.87}), the vertex corrections for the current--current
correlation functions entering the expressions for the transport
coefficients are finite when non--resonant scattering is included. The
resulting expression for the transport time after inclusion of vertex
corrections is
\begin{equation}
\frac{1}{\tau(\omega,T)}
=
\frac{1}{\tau_{0}(\omega,T)}
\left[\cos{2\eta_1} -
\frac{Re \;G_{0}(\omega,T)}{Im \;G_{0}(\omega,T)}\sin{2\eta_1}
\right]
+ \rho_n.
\label{eq:tau-multi-channel}
\end{equation}
The effects of non--resonant scattering are primarily contained in the
factor $\left[\cos{2\eta_1} -\frac{Re \;G_{0}(\omega,T)} {Im
\;G_{0}(\omega,T)}\sin{2\eta_1}\right]$ which is due to interference between
the
$l=0$ and $l=1$ channels. The non--resonant, $l\ne 0$, phase--shifts
in (\ref{eq:tau-multi-channel}) are taken to be constants, defined by
the screening charge in respective channels. Thus,
\begin{equation}
\rho_n=
{{4\pi}\over mk_F}
\left[
\sin^{2}{\eta_1}
+
\sum_{l>1}
l\;{\sin^2(\eta_l-\eta_{l-1})}
\right].
\label{eq:rho-n}
\end{equation}

In the $T=0$ limit (but with $l \neq 0$) the transport coefficients
calculated with (\ref{eq:thermo}) -- (\ref{eq:rho-n}) reduce to
standard phase--shift expressions, while in the limit $\eta_{l\neq 0}
\longrightarrow 0$ we recover the usual many--body expression for the
single--channel transport time (\ref{eq:tau-resonant-only}).

 From the expressions for $\tau(\omega,T)$ and $\tau_{0}(\omega,T)$ we
see that in order to evaluate the transport coefficients we require an
accurate expression for the frequency and temperature dependent
resonant Green's function $G_{0}(\omega,T)$. We obtain this from the
numerical renormalization group method as described below.

\subsection*{The numerical renormalization group method}

The numerical renormalization group for the Kondo and Anderson
impurity models is described in \cite{wilson.75,krishnamurthy.80},
where it was used to obtain the thermodynamic properties.  Here we
give a description of the method and its use in calculating finite
temperature Green's functions and specifically the local Green's
function $G_{0}(\omega,T)$ required for the transport time. The
central idea in the numerical renormalization group is the importance
of including all energy or length scales.  The Hamiltonian
(\ref{eq:model}) contains conduction electron states of all energies
from the band edge $D$ down to zero energy and states from each energy
scale contribute to the impurity properties.  To take into account
these states Wilson introduced a logarithmic discretization of the
conduction band about the Fermi level so that all energy scales were
represented, with the greatest resolution at low energies where the
many body effects are most important. As shown in Appendix~2 this
logarithmic discretization approximation results in the following
discrete Anderson model for the resonant channel,

\begin{eqnarray}
H & = & lim_{N\rightarrow \infty}{1\over 2}(1+\Lambda^{-1})
\Lambda^{-(N-1)/2}H_{N},\nonumber\\
H_{N} & = & \Lambda^{(N-1)/2}[ H_{0} + H_{hyb} +
\sum_{n=0}^{N-1}\Lambda^{-n/2}\xi_{n}(f_{n+1\sigma}^{\dagger}
f_{n\sigma}+f_{n\sigma}^{\dagger}f_{n+1\sigma})],
\label{eq:discrete-form}
\end{eqnarray}
where $H_{0}=\epsilon_{0}c_{0\sigma}^{\dagger}c_{0\sigma} +
Un_{0\uparrow}n_{0\downarrow}$ is the resonant part of the impurity,
$H_{hyb}=V_{0}(f_{0\sigma}^{\dagger}c_{0\sigma} + H.c.)$ couples the
impurity to a local conduction electron orbital
$f_{0\sigma}^{\dagger}|0>$, and the last term describes the remaining
conduction electron orbitals whose wavefunctions have a large overlap
with the impurity. The conduction electron orbitals neglected in the
above discrete approximation to the full Anderson model have their
wavefunctions localized away from the impurity site and have
negligible contributions to the impurity properties (see Appendix~1
and
\cite{krishnamurthy.80} for further details). The parameter $\Lambda>1$
describes the discretization of the conduction band. The above
discretized Hamiltonian is in the form of a semi--infinite linear
chain and can be iteratively diagonalized for increasing chain lengths
$N$ to obtain the eigenvalues, $E_{p}^{N}$, and eigenvectors,
$|p>_{N}$ , on successively lower energy scales $\omega_{N}\approx
D\Lambda^{-\frac{N-1}{2}}$, where $\omega_{N}$ is the lowest scale of
$H_{N}$ (see Appendix~2 for details). From the eigenvalues, the
thermodynamic properties are obtained at a corresponding sequence of
temperatures $T_{N}\approx \omega_{N}/k_{B}$.  The details are given
in \cite{krishnamurthy.80} and in Appendix~2. Here we show how the
local Green's functions can be extracted on successively lower energy
scales.

Consider the Green's function $G_{0}(\omega,T)=<<c_{0\sigma};
c_{0\sigma}^{\dagger}>>$. Using $H_{N}$, the N'th shell Green's
function $G_{0}^{N}(\omega,T)$ and associated spectral density
$\rho_{0}^{N}(\omega,T)$ are given by
\begin{equation}
G_{0\sigma}^{N}(\omega,T) = <<c_{0\sigma};c_{0\sigma}^{\dagger}>>
={1\over{Z_{N}(\beta)}}\sum_{pp'}{{{|M_{pp'}^N|}^2{(\thinspace
e^{-\beta E_{p}^N}+e^{-\beta E_{p'}^{N}}\thinspace)}}\over{\omega -
E_{p'}^N + E_{p}^N}},
\end{equation}
\begin{equation}
\rho_{0}^{N}(\omega,T)  = {1\over{Z_{N}(\beta)}}\sum_{pp'}{{{|M_{pp'}^N|}^2
(\thinspace e^{-\beta E_{p}^N}+e^{-\beta
E_{p'}^{N}}\thinspace)\delta(\omega - E_{p'}^N + E_{p}^N}}).
\label{eq:spectral-density}
\end{equation}
Here $Z_{N}(\beta)$ is the partition function for the $N'th$ cluster,
and $M_{pp'}^{N}=<p|c_{0\sigma}|p'>$ are the many--body matrix
elements of the local operator $c_{0\sigma}$.  The latter can be
evaluated recursively in a similar way to the evaluation of the matrix
elements $<p|f_{N\sigma}|q>$ in (\ref{eq:f-matrix-elements}). Using
the unitary transformation (\ref{eq:unitary-trans}) we obtain
\begin{eqnarray}
M_{pp'}^{N} & = & <p|c_{0\sigma}|p'>_{N} =
\sum_{r,i}\sum_{r',i'}U_{N}^{*}(p,ri)U_{N}(q,r'i')
<i|<r|c_{0\sigma}|r'>|i'>\nonumber\\ & = &
\sum_{r,i}\sum_{r',i'}U_{N}^{*}(p,ri)U_{N}(q,r'i')
\delta_{i,i'}M_{r,r'}^{N-1}.
\label{eq:m-matrix-elements}
\end{eqnarray}
Hence the matrix elements $M_{pp'}^{N}$ can be evaluated recursively
from a knowledge of the eigenstates of the N'th cluster,
$U_{N}(p,ri)$, and the matrix elements, $M_{rr'}^{N-1}$, of the
previous cluster.  For a given cluster size $N$ the Hamiltonian
$H_{N}$ only describes excitations in a limited range of width
$K\omega_{N}$ above the lowest scale $\omega_{N}$ present in $H_{N}$,
due to the truncation of the spectrum as described in Appendix~1.  At
$T=0$ the spectral density is evaluated at $\omega \approx
2\omega_{N}$. Calculating the spectral density at energies much
smaller than this using $H_{N}$ is not justified, since information on
lower energies is obtained in subsequent iterations, whereas
calculating the spectral densities at much higher energies than this
might introduce errors due to the truncation of the spectrum on the
high energy side. In evaluating (\ref{eq:spectral-density}) the delta
functions are broadened with Gaussians of width $\alpha_{N}$ of order
$\omega_{N}$ appropriate to the cluster size. The broadening parameter
$\alpha_{N}$ used within each energy shell is varied continuously so
that there is no discontinuity in going between successive iterations.
The small remaining asymmetry in the spectral features due to the
larger broadening parameter at the higher excitations should vanish in
the limit $\Lambda \rightarrow 1$. The procedure for calculating
finite temperature Green's functions is slightly more complicated. The
shell Green's function $G_{0}^{N}(\omega,T)$ and spectral density
$\rho_{0}^{N}(\omega,T)$ are only reliable for excitations $\omega
\approx 2\omega_{N}$ and for temperatures $k_{B}T << 2\omega_{N}$. For
temperatures $k_{B}T >> 2\omega_{N}$ excited states not contained in
$H_{N}$ would be important, whilst for temperatures $k_{B}T$ of order
$2\omega_{N}$ transitions between {\em excited} states would need to
be known accurately. In principle these are known from subsequent
iterations, but are not contained with sufficient accuracy in $H_{N}$.
The only transitions known with sufficient accuracy in $H_{N}$ are the
groundstate to excited state transitions with excitation energies
$\approx 2\omega_{N}$ which is the natural energy scale of this {\em
cluster}\footnote{The term cluster is suggested by the notation
although the calculations are in k--space}. As long as $k_{B}T <<
2\omega_{N}$ it is not necessary to know the lower excitations since
their contribution to the Green's function and spectral density {\em
for the energies $\omega = 2
\omega_{N}$ under consideration} will be negligible (note the delta function in
(\ref{eq:spectral-density})). From this discussion it follows that the
spectral density for temperature $T$ can be calculated from the shell
spectral densities $\rho_{0}^{i}(\omega,T)$ at frequencies $\omega
\approx 2\omega_{i}$ for $i=1,2,\ldots,M$ until $2\omega_{M}$ becomes
of order $T$. To calculate the spectral density at temperature $T$ and
for frequencies below $2\omega_{M}$, a smaller cluster is used. This
is done because when $T$ is larger than the frequency at which the
spectral density is being evaluated, it is the excited states of order
$T$ contained in previous clusters which are important and not the
excitations very much below $T$. The procedure outlined here requires
storing all the matrix elements for the Green's functions for each
cluster size, since smaller clusters may be required in subsequent
iterations.

In the absence of non--resonant scattering we can calculate the
transport time $\tau_{0}(\omega,T)$ directly from the single--particle
spectral density $\rho_{0}(\omega,T)$ using
(\ref{eq:tau-resonant-only}) and hence the transport coefficients
(\ref{eq:resistivity}--\ref{eq:hall-coefficient}). In the presence of
non--resonant scattering, we evaluate in addition the real part of
$G_{0}(\omega,T)$ and use (\ref{eq:tau-multi-channel}) and
(\ref{eq:resistivity}--\ref{eq:hall-coefficient}) to calculate the
transport coefficients. In the next section we present the numerical
results obtained using this procedure.

\subsection*{Low temperature results}

Before presenting numerical results we outline some analytic results
on the low temperature behaviour of the transport coefficients
obtained by using Fermi liquid theory. These serve as an independent
check on the accuracy of the numerical renormalization group method.
In the following analytic calculations we restrict ourselves to the
case of resonant scattering only so the transport time
$\tau(\omega,T)=\tau_{0}(\omega,T)$.

To extract the low temperature behaviour of the transport coefficients
we use the Sommerfeld expansion. In the transport integrals
(\ref{eq:transport-integrals}) the factor $\left(-{{\partial f}\over
{\partial\omega}}\right)$ for temperature $T$ provides an energy cut
off outside the Fermi window $|\omega |<T$. The functional form of
$G_0(\omega,T)$ also has an energy scale $|\omega|<<k_BT_0$ over which
$\tau(\omega,T)$ is a a slowly varying function, where $T_{0}=T_{K}$
for the Kondo regime and $k_BT_{0}=\Delta$, for the mixed valent
regime. As a consequence, for temperatures $T<<T_0$, we can apply the
Sommerfeld expansion,
\begin{eqnarray}
\int_{-\infty}^{+\infty}\left(-\frac{\partial f}
{\partial\omega} \right)h(\omega,T)d\omega\nonumber\\ & = &
h(\epsilon_{F},T) +
\frac{\pi^2}{6}(k_{B}T)^{2}\left(\frac{\partial^{2}h(\omega,T)}
{\partial\omega^{2}} \right)_{\omega=\epsilon_{F}}\nonumber\\ & +
&\frac{7\pi^4}{720}(k_{B}T)^{4}\left(\frac{\partial^{4}h(\omega,T)}{\partial\omega^{4}}
\right)_{\omega=\epsilon_{F}}+O(T^6).\label{eq:sommerfeld-expansion}
\end{eqnarray}
The quantity $h(\omega,T)$ entering the above expression is
$\tau^{l}(\omega,T)\omega^{m}$ for the integral $L_{ml}$.  For the
Kondo problem the transport time $\tau(\omega,T)$ is a strong function
of temperature at the Fermi level, so in the Sommerfeld expansion in
addition to the temperature dependence coming from the fermi function
there is also the explicit temperature dependence of the transport
time at the Fermi level.  This is taken into account by expanding
$\tau(\epsilon_{F},T)$ in a Taylor series
\begin{equation}
\tau(\epsilon_{F},T) = \tau(\epsilon_{F},0) + \frac{1}{2}T^{2}\left(
\frac{\partial^{2}\tau(\epsilon_{F},T)}{\partial T^{2}}\right)_{T=0} +
O(T^4)\label{eq:taylor}
\end{equation}

In the Sommerfeld expansions for the resistivity, thermopower, thermal
conductivity and Hall coefficient the derivatives which enter are
$\frac{1}{\tau}\frac{\partial\tau}{\partial\omega}$,
$\frac{1}{\tau}\frac{\partial^{2}\tau}{\partial\omega^{2}}$, and
$\frac{1}{\tau}\frac{\partial^{2}\tau}{\partial T^{2}}$. Since
$\tau_{0}(\omega,T) = 1/\Delta\rho_{0}(\omega,T)$ these derivatives
are given by

\begin{eqnarray}
\frac{1}{\tau}\frac{\partial\tau}{\partial\omega} & = &
-\frac{1}{\rho_{0}}\frac{\partial\rho_{0}}{\partial\omega},\label{eq:deriv1}\\
\frac{1}{\tau}\frac{\partial^{2}\tau}{\partial\omega^{2}} & = &
2\left(\frac{1}{\rho_{0}}\frac{\partial\rho_{0}}{\partial\omega}\right)^{2}
-\left(\frac{1}{\rho_{0}}\frac{\partial^{2}\rho_{0}}
{\partial\omega^{2}}\right),\label{eq:deriv2}\\
\frac{1}{\tau}\frac{\partial^{2}\tau}{\partial T^{2}} & = &
-\left(\frac{1}{\rho_{0}}\frac{\partial^{2}\rho_{0}}{\partial
T^{2}}.\right)
\label{eq:deriv3}
\end{eqnarray}
Using (\ref{eq:spec-density}) we obtain for the energy derivatives at
$\omega=\epsilon_{F}$ and $T=0$,

\begin{eqnarray}
\frac{1}{\rho_{0}}\left(\frac{\partial\rho_{0}}{\partial\omega}\right)
 & = & 3\gamma\cot(\pi n_{0}/2),\label{eq:first-energy-derivative}\\
\left(\frac{1}{\rho_{0}}\frac{\partial^{2}\rho_{0}}{\partial\omega^{2}}
\right) & = &
-\frac{1}{\Delta}\left(\frac{\partial^{2}\Sigma^{I}}
{\partial\omega^{2}}\right)+
\left( \frac{2\pi\rho_{0}(\epsilon_{F},0)}{z}\right)^{2} \cot^{2}(\pi n_{0}/2)
\nonumber\\
& + &
2\pi\Delta\rho_{0}(\epsilon_{F},0)\left(-\frac{1}{\tilde{\Delta}^{2}}
+\frac{1}{\Delta}\left(\frac{\partial^{2}\Sigma^{I}}
{\partial\omega^{2}}
\right)-\frac{\cot(\pi n_{0}/2)}{\Delta}\left(\frac{\partial^{2}\Sigma^{R}}
{\partial\omega^{2}}\right)
\right) ,\label{eq:second-energy-derivative}
\end{eqnarray}
where $\tilde{\Delta}=z\Delta$ and $z=(1-{{\partial \Sigma^{R}}\over
{\partial
\omega}})^{-1}_{\omega=\epsilon_{F}}$ is the wavefunction renormalization
constant.  In deriving the above we have used (\ref{eq:spec-density})
and the Fermi liquid properties of the self--energy
$\Sigma^{R}(\omega,0)\sim \omega$ and $\Sigma^{I}(\omega,0)\sim
(\omega-\epsilon_{F})^{2}$. The quantity $n_{0}$ is the local level
occupancy and $\gamma$ is the linear coefficient of the low
temperature specific heat. The latter is given by ~\cite{hewson.93}
\begin{equation}
\gamma = \frac{2}{3}\pi k_{B}^{2}\frac{\tilde{\Delta}}
{{\tilde\epsilon_{0}}^{2}+{\tilde{\Delta}}^{2}},
\label{eq:linear-coefficient}
\end{equation}
where
$\tilde{\epsilon_{0}}=z(\epsilon_{0}+\Sigma^{R}(\epsilon_{F},0))$. In
the Kondo regime $z<<1$ and $\tilde{\Delta} \sim k_BT_{K}<<\Delta$
which leads to a large enhancement of $\gamma$ since from
(\ref{eq:linear-coefficient}) $\gamma\sim 1/T_{K}$.  The Kondo
temperature, $T_{K}$, is given by
\cite{haldane.78}

\begin{equation}
k_{B}T_{K} = U{\left(\Delta\over 2U \right )^{1/2}}e^{\pi
\epsilon_{0}(\epsilon_{0}+U)/{2\Delta U}}\label{eq:kondo-temp}
\end{equation}

For the thermopower we require only the first energy derivative of
$\rho_{0}$ to extract the leading term.  We have from
(\ref{eq:thermo}) and (\ref{eq:sommerfeld-expansion}),

\begin{eqnarray}
S(T) & = & -\frac{1}{|e|T}\frac{\frac{\pi^2}{6}(k_{B}T)^{2}
\left[\frac{\partial^{2}[\omega\tau(\omega)]}
{\partial\omega^{2}}\right]_{\omega=\epsilon_{F}} + O(T^4)}
{\tau(\epsilon_{F})\left[1 +
\frac{\pi^2}{6}(k_{B}T)^{2}\frac{1}{\tau(\epsilon_{F})}
\left[\frac{\partial^{2}\tau(\omega)}
{\partial\omega^{2}}\right]_{\omega=\epsilon_{F}}\right]+O(T^4)}\nonumber\\
\label{eq:sommerfeld-thermo}
\end{eqnarray}
which on using (\ref{eq:deriv1},\ref{eq:first-energy-derivative})
reduces to the Fermi liquid result, \cite{houghton.87,kawakami.87}
\begin{equation}
S(T) = \frac{\pi\gamma T}{|e|}\cot(\pi
n_{0}/2)+O(T^{3}),\label{eq:thermo-fliq}
\end{equation}
and can be checked by evaluating $S$, $\gamma$ and $n_{0}$ within the
numerical renormalization group technique \cite{costi.93a}.

To extract the other transport coefficients to lowest order in $T$
requires the second derivative with respect temperature of $\tau_{0}$
and hence an expression for the self--energy correct to order $T^{2}$.
This is difficult to obtain analytically in the general case so we
restrict ourselves to just the symmetric case. The self--energy for
the symmetric model correct to order $\omega^2$ and $T^2$ is given by
\cite{yamada.75a,yamada.75b}
\begin{equation}
\Sigma(\omega,T) = \Sigma^{R}(\omega,T) + i\Sigma^{I}(\omega,T) =
\Sigma^{R}(\omega,0) - i\frac{\Gamma_{0}^{2}}{2\Delta(\pi\Delta)^{2}}(\omega^2
+ \pi^{2}(k_{B}T)^{2}),\label{eq:self-energy-order2}
\end{equation}
where $\Gamma_{0}$ is the irreducible vertex function evaluated at
zero frequency \cite{yamada.75a,yamada.75b}.  Using
(\ref{eq:self-energy-order2}) we obtain the temperature derivative,
\begin{eqnarray}
\left(\frac{1}{\tau}\frac{\partial^{2}\tau}{\partial T^{2}}
\right)_{\omega=\epsilon_{F},T=0} & = &
-\left(\frac{1}{\rho_{0}}\frac{\partial^{2}\rho_{0}}{\partial
T^{2}}\right) =
\frac{k_{B}^{2}\Gamma_{0}^{2}}{\pi\Delta}\frac{1}{ ((\epsilon_{0}+\Sigma^{R})^2
+ \Delta^{2})^2}.\label{eq:second-temp-derivative}
\end{eqnarray}
We now use these results to first check the exact result for the
$T^{2}$ coefficient of the resistivity first derived by Nozi\`eres
\cite{nozieres.74} and Yamada \cite{yamada.75a,yamada.75b} and then
derive the exact coefficient for the $T^2$ coefficient of
$(\kappa(T)/T)/(\kappa(T)/T)_{T\rightarrow 0}$.  Restricting the
discussion to the Kondo regime where the charge susceptibility is zero
so $\Gamma_{0}=\pi\Delta/z$, and the particle-hole symmetric model
where $\epsilon_0+\Sigma^R=0$, we find from
(\ref{eq:second-energy-derivative},
\ref{eq:second-temp-derivative})
\begin{equation}
\left(\frac{1}{\tau}\frac{\partial^{2}\tau}{\partial\omega^{2}}\right)
_{\omega=\epsilon_{F},T=0}= 3/\tilde{\Delta}^{2}, \;\;\mbox{and}\;\;
\left(\frac{1}{\tau}\frac{\partial^{2}\tau}{\partial T^{2}}
\right)_{\omega=\epsilon_{F},T=0}=\pi^{2}k_{B}^{2}/
\tilde{\Delta}^{2}.\label{eq:derivs}
\end{equation}
Hence from
(\ref{eq:resistivity},\ref{eq:sommerfeld-expansion},\ref{eq:taylor})
and (\ref{eq:derivs}) we have for the resistivity,
\begin{eqnarray}
\rho(T) & = &\frac{1}{e^{2}\tau(\epsilon_{F},0)\left[1 + \frac{1}{2}T^{2}
\frac{1}{\tau(\epsilon_{F},0)}\frac{\partial^{2}\tau}{\partial T^{2}} +
\frac{\pi^{2}(k_{B}T)^{2}}{6}
\frac{1}{\tau(\epsilon_{F},0)}\frac{\partial^{2}\tau}{\partial \omega^{2}}
+ O(T^4)\right]},\nonumber\\ & = &
\frac{1}{e^{2}\tau(\epsilon_{F},0)\left[1 +
\frac{1}{2}
\frac{\pi^2 k_{B}^2}{\tilde{\Delta}^2}T^{2} + \frac{1}{2}\frac{\pi^2 k_{B}^2}
{\tilde{\Delta}^2}T^{2}+ O(T^4)\right]},\nonumber\\ & = & \rho(0)(1 -
c
\left(\frac{T}{T_{K}}\right)^{2}),
\end{eqnarray}
where $\rho(0)=1/e^{2}\tau(\epsilon_{F},0)$ and
$c=\pi^{2}k_{B}^{2}T_{K}^2/\tilde{\Delta}^2=\pi^4/16=6.088$ and we
have used the result for the symmetric case $\pi\tilde{\Delta}=4
k_{B}T_{K}$
\cite{yamada.75a,yamada.75b}.  The above result for $c$ is the exact result
obtained by Nozi\`eres \cite{nozieres.74}. Without the contribution
from (\ref{eq:second-temp-derivative}) the coefficient $c$ would have
been wrong by a factor of 2. Away from the symmetric case the result
of Nozi\`eres will be modified due terms in
(\ref{eq:deriv1}--\ref{eq:deriv2}) which depend on the occupation
number $n_{0}$ and also due to additional temperature dependences
coming from the occupation number and the imaginary part of the
self--energy.  These are difficult to estimate analytically, but our
numerical results to be presented in the next section give some
indication of their size. Most terms depend on $\cot(\pi n_{0}/2)$ and
are expected to be small in the Kondo regime $n_{0}\approx 1$.

We now prove an exact Fermi liquid result for the thermal
conductivity. Using $|e|S(T)/T = \pi\gamma \cot(\pi n_{0}/2)$ we have

\begin{equation}
\frac{\kappa(T)}{T}=\frac{1}{T^2}L_{21}+\pi\gamma \cot(\pi n_{0}/2)L_{11}.
\end{equation}

The expansion for $L_{21}$ is
\begin{eqnarray}
L_{21} & = & \int_{-\infty}^{+\infty}\left(-\frac{\partial f}
{\partial\omega}
\right)\omega^{2}\tau(\omega,T)d\omega,\nonumber\\ & = &
\frac{\pi^2}{6}(k_{B}T)^{2}\left(\frac{\partial^{2}(\omega^{2}\tau(\omega))}
{\partial\omega^{2}} \right)_{\omega=\epsilon_{F}} +
\frac{7\pi^4}{720}(k_{B}T)^{4}\left(\frac{\partial^{4}(\omega^{2}\tau(\omega))}
{\partial\omega^{4}} \right)_{\omega=\epsilon_{F}}+O(T^6),\nonumber\\
& = &
\frac{\pi^2 k_{B}^2 T^{2}}{3}\tau(\epsilon_{F},0)
\left[ 1 + \frac{3}{2\pi^2 k_{B}^2}T^{2}
\frac{1}{\tau}\left(\frac{\partial^2 \tau}{\partial T^2}
\right)_{T=0}\right] + \frac{7\pi^4}{60}(k_{B}T)^4
\left(\frac{\partial^2 \tau}
{\partial \omega^{2}}\right)_{\omega=\epsilon_{F}}+O(T^6),\nonumber\\
& = &
\frac{\pi^2 k_{B}^2 T^{2}}{3}\tau(\epsilon_{F},0)
\left[ 1 + \frac{3}{2\pi^2 k_{B}^2}T^{2}
\frac{1}{\tau}\left(\frac{\partial^2 \tau}{\partial T^2}
\right)_{T=0}
 + \frac{7\pi^2 k_{B}^{2}T^2 }{20}\frac{1}{\tau}
\left(\frac{\partial^2 \tau}
{\partial \omega^{2}}\right)_{\omega=\epsilon_{F}}+O(T^4)\right].
\label{eq:l21}
\end{eqnarray}
Substituting this into the expression for the thermal conductivity,
and using the above results for the derivatives we obtain for the
symmetric case in the Kondo limit,
\begin{equation}
\frac{\kappa(T)}{T}/(\frac{\kappa(T)}{T})_{0} = 1 +
\beta \left(\frac{T}{T_{K}}\right)^{2},
\end{equation}
where $\beta=26c/10=13\pi^{4}/80=15.83$ and the linear coefficient of
$\kappa$ is $\alpha = \left(\frac{\kappa(T)}{T}\right)_{0}=
\pi^{2}k_{B}^{2}\tau(\epsilon_{F},0)/3$. In the asymmetric case there will be
additional contributions
to the thermal conductivity which will modify the coefficient $\beta$.
As discussed earlier for the resistivity these will be difficult to
estimate but are expected to be small in the Kondo regime, so we
expect the coefficient $\beta$ to remain close to its symmetric value
$13\pi^{4}/80$.

The low temperature behaviour of the Hall coefficient can also be
extracted.  From the Sommerfeld expansion it can be shown that the
order $T^{2}$ terms from the energy derivatives cancel and that the
$T^{2}$ term results from the temperature dependence of $\tau$ at the
Fermi level, yielding in the symmetric case $R_{H}(T)\sim
-R_{\infty}(1-\delta(T/T_{K})^{2})$ where the coefficient $\delta$
depends on second derivatives and a fourth derivative $(\partial^4
\tau/\partial \omega^{2}\partial T^{2})_{\omega=\epsilon_{F},T=0}$.  We do not
evaluate this since as discussed in the section on numerical results,
there is a much larger contribution to the Hall effect due to skew
scattering of electrons which should be included for a discussion of
the Hall coefficient of magnetic impurities \cite{coleman.85,fert.73}.

To summarize the analytic calculations, we see that in Fermi liquid
theory in the Kondo regime that the transport coefficients all scale
with $T_{K}$ at low temperature.  Specifically we have

\begin{equation}
\rho(T) = \rho(0)(1-c\thinspace (T/T_{K})^{2}),\label{eq:scaling1}
\end{equation}
\begin{equation}
S(T) = \frac{\pi\gamma T}{|e|}\cot(\eta_{0}(\epsilon_{F}))\sim
(T/T_{K}),\label{eq:scaling2}
\end{equation}
\begin{equation}
\left(\kappa(T)/T\right)/\left(\kappa(T)/T\right)_{0} =
1 + \beta (T/T_{K})^{2},\label{eq:scaling3}
\end{equation}
\begin{equation}
R_{H}(T)/R_{\infty} = -(1-\delta (T/T_{K})^{2}),\label{eq:scaling4}
\end{equation}
where the constants $c,\alpha,
\beta,\gamma$ and $\delta$ depend on $\tau$ and its derivatives at the Fermi
level.
For the symmetric model one finds,
\begin{eqnarray}
c & = & \frac{\pi^{4}}{16} = 6.088,\label{eq:constants1}\\
\alpha & = & (\frac{\kappa(T)}{T})_{0} =
\frac{k_{B}^{2}\pi^{2}}{3}\tau(\epsilon_{F},0),\label{eq:constants2}\\
\beta  & = & (13/80)\pi^{4} = 15.83,\label{eq:cosntants3}\\
\gamma & = &\frac{\pi^2}{6k_{B}T_{K}}.\label{eq:constants4}
\end{eqnarray}
Corresponding expression for arbitrary $U$ can be derived from a
renormalized perturbation theory in terms of $\tilde \Delta$ and a
renormalized interaction $\tilde U=z^2\Gamma_0$ \cite{rpt}.\par In the
asymmetric case for the Kondo regime $n_{0}\approx 1$, there should be
negligible corrections to these coefficients, the asymmetry in this
regime should only affect the value of $T_{K}$.

The functional form of the low temperature transport coefficients in
the Kondo regime (\ref{eq:scaling1}--\ref{eq:scaling4}) and the
coefficients of the leading order terms
(\ref{eq:constants1}--\ref{eq:constants4}) in the symmetric case
provide a check on the accuracy of the numerical renormalization group
results presented in the next section.

\section*{Numerical results}

The numerical results presented here are for the strong correlation
limit of the Anderson model, i.e. $U/\pi\Delta >> 1$. We take
$U/\pi\Delta=4$, unless otherwise indicated. The local level position
takes values $\epsilon_{0}/\Delta=-U/2\Delta=-2\pi$ corresponding to
the symmetric case and $\epsilon_{0}/\Delta=-4,-3,-2,-1,0,+1$
corresponding to the asymmetric case. We are mainly interested in
universal effects independent of the band structure, so all parameters
are small relative to the conduction electron half--bandwidth $D=1$.
Specifically we used $\Delta=0.01D$.  For
$\epsilon_{0}/\Delta=-4,-3,-2$ the parameters describe the Kondo
regime, whilst for $\epsilon_{0}/\Delta=-1,0$ and
$\epsilon_{0}/\Delta=+1$ they describe mixed valent and empty orbital
regimes respectively. In discussing the effects of non--resonant
scattering on the transport properties we took a small $l=1$ phase
shift, $\eta_{1}=\pm 0.01\pi,\pm 0.02\pi,\ldots,\pm 0.05\pi$. The
discretization parameter $\Lambda$ can be taken as low as $1.5$
without encountering large errors due to the truncation of high energy
states. The use of such a small value of $\Lambda$ does introduce
errors into the spectral densities for extremely low energies.  These
are noticeable in the symmetric case for example, where for
sufficiently low energies $\omega << k_{B}T_{K}$, the symmetry of the
spectral density is violated slightly. A larger value of $\Lambda$
allowing the lower energies to be reached in fewer iterations could be
used to avoid this problem, however this was not necessary in the
present calculations.

\subsection*{Thermodynamics}

The thermodynamic properties of the Anderson model have been discussed
in
\cite{krishnamurthy.80} where the static susceptibility was calculated over the
whole temperature range from the band edge down to temperatures $T <<
T_{K}$.  The specific heat was calculated in the strong coupling
regime $T<<T_{K}$ by perturbation theory on the effective Hamiltonian
around the strong coupling fixed point \cite{krishnamurthy.80}.  Here
we briefly present the results for the local level occupancy
$n_{0}(T)$ and specific heat $C(T)$ over the whole temperature regime
of interest (calculations of $C(T)$ for the screened Anderson model
where presented in \cite{costi.91}).

\subsubsection*{Local level occupancy, $n_{0}(T)$}

The local level occupancy $n_{0}(T)$ can be calculated at a sequence
of temperatures $T_{N}, N=1,2,\ldots$ from the average electron number
on the impurity and is shown in Fig.\ \ref{fig-valence} for several
values of the local level position ranging from the Kondo regime
($\epsilon_{0}/\Delta = -4,-3,-2$) to the mixed valent
($\epsilon_{0}/\Delta = -1,0$) and empty orbital regime
($\epsilon_{0}/\Delta = +1$). The only significant temperature
variation occurs on the valence fluctuation temperature scale $k_{B}T
= \Delta=10^{-2}D$.  On the Kondo temperture scale, $T_{K}$ (listed in
Table \ref{table1}), the variation of $n_{0}$ is insignificant
indicating that the Kondo resonance has negligible weight in agreement
with the spectral densities presented below.  The values of
$n_{0}(T=0)$ are shown in Table \ref{table1}.

\subsubsection*{Specific heat}

The specific heat, shown in Fig.\ \ref{fig-specific-heat-all},
exhibits one or two distinct peaks depending on the position of the
local level. In the Kondo regime there are two distinct peaks, one at
$T\approx T_{K}$ and another at $T$ of the order of $\Delta$.  The low
energy peak, which is observed in many systems, is associated with the
entropy of the magnetic impurity so the area under this peak is given
by $k_{B}ln(2)$ for the present case where the impurity has $S=1/2$.
The higher peak is associated with the charge degrees of freedom.  The
same behaviour is found in the Bethe--Ansatz solution \cite{okiji.83}.
In Fig.\ \ref{fig-specific-heat-gamma} we show the ratio $C(T)/T$.
The linear in $T$ behaviour characteristic of a Fermi liquid is clear
and we tabulate the linear coefficient in Table \ref{table1}.
Finally, Fig.\
\ref{fig-specific-heat-universal} shows that the specific heat is a universal
function of $T/T_{K}$ in the Kondo regime. Universality for this
quantity extends up to $T\approx 5T_{K}$ for the case
$\epsilon_{0}/\Delta = -4$. From the figure it seems very likely that
in the Kondo regime and provided the local level is very far from the
Fermi level the universal behaviour should extend over the whole range
$10T_{K}$ of the low low energy peak.

In the mixed valent regime there is a peak slightly below the valence
fluctuation temperature scale $k_{B}T=\Delta$ and a shoulder or small
peak at higher temperatures. The same behaviour is found in the empty
orbital case. The same behaviour is found also from the Bethe--Ansatz
solution \cite{okiji.83}.

The value of the linear coefficient of specific heat has been checked
in the case of $U=0$ and found to be in good agreement with the exact
result. It has also been checked independently for both $U=0$ and
finite $U$ \cite{costi.93c} by using the Fermi liquid relation,
\begin{equation}
{4\chi_{s} \over (g\mu_{B})^{2}} + \chi_{c} = {6 \over
{(k_{B}\pi)^{2}}}\gamma,
\end{equation}
where $\chi_{s}$ and $\chi_{c}$ are the spin and charge
susceptibilities and can be obtained as thermodynamic quantities or
from the corresponding dynamic susceptibilities. This Fermi liquid
relation has been shown to be satisfied within a few percent and
provides an independent test on the accuracy of $\gamma$ from the
present thermodynamic calculation \cite{costi.93c}

\subsection*{Spectral densities}

%
%
The spectral densities in the Kondo regime are shown in Fig.\
\ref{fig-spec-zero-kondo}--\ref{fig-spec-finite-asym-kondo} for zero and finite
temperatures.  At low temperatures $T<< T_{K}$ the spectral density is
characterized by three resonances: the Kondo resonance at the Fermi
level of width $T_{K}$ and negligible weight and two atomic--like
resonances on either side close to the bare excitation energies
$\epsilon_{0}$ and $\epsilon_{0}+U$.  The latter contain most of the
spectral weight and have widths of order $\Delta$. The Kondo resonance
lies at the Fermi level in the symmetric case $\epsilon_{0}=-U/2$ and
above the Fermi level in the asymmetric case $\epsilon_{0}>-U/2$ (see
inset to Fig.\ \ref{fig-spec-zero-kondo}). The width of the Kondo
resonance is approximately $T_{K}$, where $T_{K}$ is defined in
(\ref{eq:kondo-temp}) and shows the correct exponential dependence on
$\epsilon_{0}$ and $U$ (inset to Fig.\ \ref{fig-spec-zero-kondo} and
\cite{costi.92b}).  On increasing the temperature, the Kondo resonance
decreases
rapidly in height and eventually disappears completely for $T>>T_{K}$
(Fig.\
\ref{fig-spec-finite-symm-kondo}--\ref{fig-spec-finite-asym-kondo}). The
atomic--like peaks remain temperature independent until $k_BT\approx
\Delta$.  For $k_BT \ge \Delta$ they acquire some temperature
dependence, and in the asymmetric case, Fig.\
\ref{fig-spec-finite-asym-kondo}, there is a transfer of spectral
weight from the lower to the upper peak with increasing temperature.
In the symmetric case, Fig.\ \ref{fig-spec-finite-symm-kondo}, no
spectral weight can be transferred but instead the two peaks broaden
with increasing temperature. The positions of these high energy peaks
changes little with temperature. Qualitatively similar results for the
temperature dependence of the spectral densities have been obtained by
perturbative methods for $U/\pi\Delta\le 2.5$ \cite{horvatic.87}.
%
%
The spectral densities in the mixed valent and empty orbital regimes
are shown in Fig.\
\ref{fig-spec-zero-mixed+empty}--\ref{fig-spec-finite-empty-orbital}
for zero and finite temperatures. In these regimes the spectral
density at low temperatures, $k_{B}T<<\Delta$, is characterized by two
resonances (see
Fig.\ref{fig-spec-zero-mixed+empty}--\ref{fig-spec-finite-empty-orbital}):
one at $\tilde{\epsilon_{0}}$ of width approximately $\Delta$ and a
much broader resonance at $\tilde{\epsilon_{0}}+U$ carrying very much
less weight. In the mixed valent regime, $\epsilon_{0}/\Delta \approx
-1$, the bare level is renormalized by the interactions to lie above
the Fermi level $\tilde{\epsilon_{0}}>0$. The bare level is also
renormalized to higher energies in the empty orbital case , i.e.
$\tilde{\epsilon_{0}}>\epsilon_{0}\ge \Delta$.  On increasing
temperature, the main effect in the mixed valence case is a strong
temperature dependence of the resonance at $\tilde{\epsilon_{0}}$ on a
scale of order $\Delta$ (see Fig.\
\ref{fig-spec-finite-mixed-valent}--\ref{fig-spec-finite-empty-orbital}). This
resonance broadens and decreases in height with increasing
temperature, but does not completely disappear at high temperature. In
addition it always remains above the Fermi level. The resonance at
$\tilde{\epsilon_{0}}+U$ remains distinct from that
$\tilde{\epsilon_{0}}$ up to at least $k_BT\approx 6\Delta$.  Similar
behaviour occurs in the empty orbital regime but the relevant energy
scale for the temperature dependence in this case is set by
$\tilde\epsilon_0$.

A good test on the accuracy of the spectral densities is provided by
the Friedel sum rule which relates the spectral density at the Fermi
level to the $T=0$ occupation number $n_{0}(T=0)$:
\begin{equation}
\rho_{0}(\omega=\epsilon_{F},T=0)={1 \over
 {\pi\Delta}}\sin^{2}(\pi n_{0}/2).\label{eq:friedel}
\end{equation}
Using $n_{0}(T=0)$ calculated from the partition function gives
agreement to within 3\% in all parameter regimes as shown in Table
\ref{table1}. The values of $n_{0}(T=0)$ calculated from the spectral
densities are within 3\% of those calculated from the partition
function, but the former are expected to be essentially exact since
they are obtained from the low energy part of the spectrum
($n_{0}(T=0)=\lim_{T_{N}\;\rightarrow 0}<n_{0}>_{T_{N}}$) where the
method gives the highest resolution. It should be noted that the value
of $n_{0}$ calculated from the spectral density is largely due to the
high energy features at $\epsilon_{0}$, where the method has less
resolution due to the logarithmic discretization of the conduction
band. However, the agreement to within 3\% between the values of
$n_{0}$ calculated by these two different ways provides additional
evidence that method can give accurate results for high energy
spectral features as well as low energy ones. The high energy features
have a small asymmetry due the broadening procedure used for the delta
functions in the spectral density which should vanish in the limit
$\Lambda \rightarrow 1$ (as described earlier).

\subsection*{Transport coefficients}

\subsubsection*{Resistivity}

Results for the resistivity are shown in Fig.\
\ref{fig-resistivity-all} for several values of the local level
position ranging from the Kondo regime ($\epsilon_{0}/\Delta=-{U \over
2\Delta}, -4, -3, -2$) to the mixed valent
($\epsilon_{0}/\Delta=-1,0$) and empty orbital
($\epsilon_{0}/\Delta=+1$) regimes. The behaviour of the resistivity
is qualitatively similar in all regimes with a monotonic increase with
decreasing temperature. At $T=0$ it reaches its limit $\rho(0)=1/e^2
\tau_{0}(\epsilon_{F},0)$.  Since $\tau_{0}\sim 1/\rho_{0}$ we have
from the Friedel sum rule for the spectral density that the $T=0$
resistivity $\rho(0)\sim \sin^{2}(\pi n_{0}/2)$. This is satisfied for
all cases shown in Fig.\ \ref{fig-resistivity-all} with the same
accuracy that the Friedel sum rule was satisfied for the spectral
density (discussed earlier and summarized in Table \ref{table1}).

In the Kondo regime at low temperature $T<<T_{K}$ the resistivity is
given by the exact Fermi liquid result
\cite{nozieres.74,yamada.75a,yamada.75b}
\begin{equation}
\rho(T) = \rho(0)\left\{ 1 - c \left( {T \over T_{K} }
\right)^{2}\right\},\label{eq:resis-nozieres}
\end{equation}
where $c = {{\pi}^4 \over 16} = 6.088$ with corrections which increase
with increasing asymmetry, as indicated in the previous section. The
rescaled resistivity in the Kondo regime is shown in Fig.\
\ref{fig-resistivity-universal+fliq}. Universal behaviour is found in the range
$0\le T \le 5T_{K}$. The inset shows the expected $T^{2}$ Fermi liquid
behaviour. In the Kondo regime the values of the $T^{2}$ coefficient
extracted from a least squares fit in the region $0\le T \le 0.1T_{K}$
are within about 8\% of the exact result $c=6.088$ (see Table
\ref{table1}). The values of $c$ increase systematically with
increasing asymmetry and we believe this is in part due to the
correction terms discussed previously.  However the increase in $c$ is
small, less than $8\%$ in going from the symmetric to the
$\epsilon_{0}=-2\Delta$ case, which indicates that the correction
terms discussed earlier are also small in the Kondo regime.

At high temperatures $T>>T_{K}$, the resistivity is well described by
the Hamann result \cite{hamann.67} for the sum over parquet diagrams,
\begin{equation}\rho(T)={ \rho(0)\over 2}\left(1-{{{\rm
ln}(T/T_{KH})}\over {[({\rm ln}(T/T_{KH})^2+\pi^2
S(S+1)]^{1/2}}}\right).\label{eq:hamann}
\end{equation}
Fig.\ \ref{fig-resistivity-hamann} shows the resistivity for the
symmetric case together with a fit of the Hamann result with
$T_{KH}=T_{K}/1.2$. The fit is very good in the range $T_{K}\le T \le
10T_{K}$ but (\ref{eq:hamann}) fails to give the correct Fermi liquid
behaviour at low temperature.

The effect of non--resonant scattering on the resistivity is shown in
Fig.\
\ref{fig-resistivity-nonres-symm} for the symmetric case. The resistivity
decreases slightly at low temperature for increasing absolute values
of the non--resonant phase shift $|\eta_{1}|$. The overall behaviour,
however, for small values of $|\eta_{1}|$ is the same as for resonant
scattering only. The decrease with increasing $|\eta_{1}|$ is due to
the induced change in the local charge $\delta n$ away from $n_{0}=1$.
Since the resistivity attains its maximum for $n_{0}=1$, any change
away from this leads to a smaller resistivity.  In the asymmetric
case, the effect of non--resonant scattering is to decrease the
resistivity for increasingly negative values of $\eta_{1}$ and to
increase it for increasingly positive values of $\eta_{1}$. This
effect is also due to the induced change in the local charge due to a
repulsive ($\eta_{1}>0$) or attractive ($\eta_{1}<0$) local potential.

To summarize, the resistivity is described accurately over the whole
range of temperatures from the logarithmically dominated region
$T>>T_{K}$ through the crossover region $T\sim T_{K}$ and into the low
temperature Fermi liquid region $T<<T_{K}$.

\subsubsection*{Thermopower}

The thermopower is shown in Fig.\
\ref{fig-S-low}--\ref{fig-thermo-mixed} for the case of resonant
scattering only. The effect of non--resonant scattering on the
thermopower is discussed below. In Fig.\ \ref{fig-S-low} we show the
low temperature Fermi liquid behaviour of the thermopower given by
(\ref{eq:thermo-fliq}), $S(T)/\gamma\thinspace T=(\pi/e)\cot(\pi
n_{0}/2)$, which relates the linear coefficients of the thermopower
and specific heat to the resonant level occupancy. By extracting
$\gamma$ from the second derivative $-\lim_{T\longrightarrow
0}{\partial^{2}F_{imp}(T)/\partial T^{2}}$ of the impurity Free energy
and $n_{0}$ from the spectral density or partition function we have
shown elsewhere \cite{costi.93a} that this Fermi liquid relation is
satisfied to a high degree of accuracy in all parameter
regimes\footnote{In the inset to Fig. 4 of \cite{costi.93a} the curves
$S(T)/T$ were labelled in the reverse order to the correct one shown
here for different local level positions}.

 From Fig.\ \ref{fig-thermo-kondo}--\ref{fig-thermo-mixed} it can be
seen that the thermopower has different characteristic behaviour in
the Kondo and mixed valent regimes. The thermopower for the symmetric
case, $\epsilon_{0}=-U/2$, is identically zero due to particle--hole
symmetry (see later for a discussion of the effect of the
non--resonant terms in this case). For the asymmetric case in the
Kondo regime ($\epsilon_{0}/\Delta=-4, -3,-2$) it exhibits a low
temperature maximum at $T\approx T_{K}/3$ and then changes sign for
$T>T_{K}$ before reaching a broad minimum at high temperature
$k_BT\approx \Delta$ (Fig.\
\ref{fig-thermo-kondo}). In the mixed valent regime
($\epsilon_{0}/\Delta=-1,0$)
there is again a maximum, at $k_BT\approx \Delta/3$, but there is no
sign change at higher temperature and there is only a shallow minimum
at $k_BT\approx
\Delta$ (Fig.\ \ref{fig-thermo-mixed}). The behaviour in the empty orbital
regime ($\epsilon_{0}/\Delta=+1$) is similar to that in the mixed
valent regime except that the shallow minimum at
$k_BT\approx\tilde{\epsilon_0}\approx\Delta$ becomes a shoulder (Fig.\
\ref{fig-thermo-mixed}).

This complicated behaviour of the thermopower has a clear
interpretation in terms of the temperature dependence of the spectral
densities. We consider first the Kondo regime. The low temperature
behaviour $T<<T_{K}$ of the thermopower obtained from the Sommerfeld
expansion $S(T)\sim \gamma \thinspace T$ shows that $S$ is positive
and large due to the strongly enhanced values of $\gamma$ in the Kondo
regime ($\gamma \sim 1/T_{K}$).  At temperatures $T$ of order $T_{K}$
for which the thermopower is determined by the Kondo resonance, we see
from (\ref{eq:thermo}) that the sign of the thermopower depends on the
slope of the spectral density at the Fermi level
$\left({\partial\rho_{0}(\omega,T)} \over
{\partial\omega}\right)_{\omega=\epsilon_{F}}$. This slope decreases
and eventually changes sign for $T>T_{K}$ as a result of the strong
temperature dependence of the Kondo resonance. This sign change at
$T\approx T_{K}$ is characteristic of the Kondo regime. As the
temperature is increased to $k_BT\approx \Delta$, charge fluctuations
become increasingly more important and there is a transfer of spectral
weight from the lower satellite peak at $\epsilon_{0}$ to the upper
one at $\epsilon_{0}+U$ (Fig.\
\ref{fig-spec-finite-asym-kondo}).  Holes are scattered increasingly more than
electrons as the temperature is increased and therefore the
thermopower increases for $k_BT>\Delta$. Hence a broad minimum arises
at $k_BT\approx
\Delta$.

The behaviour of the thermopower in the mixed valent and empty orbital
regimes can be explained in a similar way in terms of the temperature
dependence of the spectral density.  In the mixed valent regime the
interactions renormalize the bare local level $\epsilon_{0}$ to lie at
$\tilde{\epsilon_{0}}$ above the Fermi level (see Fig.\
\ref{fig-spec-zero-mixed+empty}).  The thermopower can be analyzed in
terms of the resonance at $\tilde{\epsilon_{0}}$ of width $\Delta$ in
a similar way to the above analysis for the Kondo resonance. The low
temperature maximum occurs at $k_BT\approx \Delta/3$ but the low
temperature enhancement $S(T)\sim \gamma T$ is only of order $\gamma
\sim 1/\Delta$ (neglecting the phase factor $\cot(\pi n_{0}/2)$) .
There is a maximum at $k_BT\approx \Delta$, beyond which the
thermopower decreases. In this case however, the resonance above the
Fermi level does not disappear but is increasingly broadened with
temperature (see Fig.\
\ref{fig-spec-finite-mixed-valent}).  Hence the thermopower does not change
sign. In the empty orbital regime, the behaviour of the thermopower
can be similarly explained in terms of the temperature dependence of
the spectral density.

We now consider the effect of non--resonant scattering of conduction
electrons on the thermopower. The effects of non--resonant scattering
are most dramatic for the symmetric case, where in the absence of such
scattering the thermopower vanishes identically.  The presence of
non--resonant scattering leads to interference effects between the
resonant $l=0$ channel and non--resonant $l=1$ channels making the
thermopower finite even for the symmetric case (see Appendix~1 for
details of how these interference terms arise). This is seen
explicitly at low temperatures by applying the Sommerfeld expansion to
(\ref{eq:thermo}) keeping the full transport time with inclusion of
non--resonant terms,

\begin{equation}
S(T) = \frac{\pi\gamma \; T}{|e|}
\left[
{\sin(2\eta_0(\epsilon_F)-2\eta_{1} )}
\over
{\sin^2(\eta_0(\epsilon_F)-\eta_1)+\rho_n}
\right] + O(T^3).
\label{eq:sommerfeld-thermo-nonres}
\end{equation}
Here, $\eta_0(\omega)$ is the resonant phase shift,
\begin{equation}
\tan[\eta_{0}(\omega)]
= -\frac{Im\;G_0(\omega)}{Re\;G_0(\omega)},
\label{eq:eta}
\end{equation}
and $\eta_{1}$ is the phase--shift of conduction electrons scattering
in the $l=1$ channel ($\rho_{n}$ contains the effects of higher
angular momentum scattering and was discussed earlier). For
$\eta_{0}(\epsilon_{F})=\pi/2$ corresponding to the symmetric case the
above expression gives a finite thermopower due to the non--resonant
terms.

In Fig.\ \ref{fig-thermo-symm} the thermopower for the symmetric case
and several values of the non--resonant phase shift $\eta_{1}$ is
shown. The thermopower can be positive or negative at low temperature
depending on whether $\eta_{1}$ is negative or positive. This is clear
since a positive $\eta_{1}$ corresponds to a repulsive potential which
decreases the charge on the impurity making the thermopower the same
as for less than half--filling, i.e. positive at low temperature and
conversely for $\eta_{1}$ negative.

The effect of non--resonant scattering of conduction electrons on the
thermopower in the asymmetric cases is shown in Fig.\
\ref{fig-thermo-kondo+nr} and Fig.\ \ref{fig-thermo-mixed+nr} for the
Kondo and mixed valent cases respectively. In the mixed valent case we
see that a small positive non--resonant phase shift can change the
minimum at $k_BT\approx \Delta$ (Fig.\
\ref{fig-thermo-mixed+nr}) into a shoulder making the thermopower resemble that
of the empty orbital case. On the other hand a small negative
non--resonant phase shift accentuates the minimum at $k_BT\approx
\Delta$ and makes the thermopower more similar to that of the Kondo
case. It is possible that a sufficiently large negative phase shift
$\eta_{1}$ could make the thermopower in the mixed valent case change
sign, but in this case, which corresponds to strong non--resonant
scattering, the frequency dependence of $\eta_{1}$ should be taken
into account. In the Kondo regime similar trends are found on varying
the non--resonant phase shift from negative to positive values. In the
weak Kondo regime, $\epsilon_{0}/\Delta = -2$ (Fig.\
\ref{fig-thermo-kondo+nr}), for which $T_{K}=0.18\Delta$, a
sufficiently large value for $\eta_{1}>0.042$ can change the sign of
the high energy minimum at $k_BT\approx \Delta$ making the thermopower
resemble that of the mixed valent case.  The effects of non--resonant
scattering can be understood as for the symmetric case as arising from
a change in the local screening charge on the impurity. A difference
in the thermopower of Fig.\ \ref{fig-thermo-symm} for the symmetric
case and Fig.\
\ref{fig-thermo-kondo+nr} for the asymmetric case is that the extremum at
$k_BT\approx \Delta$ in the symmetric case is very much suppressed.
This is expected since this extremum is associated with the charge
fluctuations which in the symmetric case are strongly suppressed and
only included in the present case as a result of the non--resonant
scattering terms. We see then that the thermopower is sensitive to
small changes in the local environment of the magnetic ion which has
been modelled here by including non--resonant scattering of conduction
electrons. The sensitivity arises because of interference between the
$l=1$ channel and the resonant $l=0$ channel, in contrast to the $l>1$
channels which have negligible influence since they do not couple to
the resonant channel.

\subsubsection*{Thermal Conductivity and Lorenz number}

The thermal conductivity is shown in Fig.\ \ref{fig-thermal-cond}.  It
shows the expected Fermi liquid behaviour $\kappa(T)/\alpha T \sim 1 +
\beta (T/T_{K})^{2}$ at low temperature ($T<<T_{K}$) with linear
coefficient $\alpha
\sim \tau(\epsilon_{F})
\sim 1/\sin^{2}(\pi n_{0}/2)$ as shown in
Fig.\ \ref{fig-thermal-cond-alpha} and in accordance with the Friedel
sum rule.  The $T^2$ coefficient $\beta$, i.e. the $T^3$ coefficient
of $\kappa(T)$, has also been extracted from the numerical results by
using a least squares fit in the region $0\le T/T_{K} \le 0.01$.  The
$T^2$ behaviour is shown in Fig.\
\ref{fig-thermal-cond-beta} and the results extracted for $\beta$ are listed in
Table \ref{table1}.  It lies within 15--25 \% of the exact result
$\beta=15.83$ calculated above using Fermi liquid theory. The
agreement is surprisingly good since we are dealing with $T^{3}$
corrections to $\kappa(T)$ and the extraction of $\beta$ also relies
on an accurate extraction of the linear coefficient $\alpha$. The
resistivity coefficient comes from a $T^{2}$ term and is consequently
more accurate.  A distinct low temperature anomaly is absent from
$\kappa(T)$, and its behaviour is similar in both the Kondo and mixed
valent regimes. Universality in $\kappa(T)/T$ extends up to at least
$10T_{K}$ in the Kondo regime. The Lorenz number ratio $L/L_{0}$ where
$L=\kappa\rho/T$ and $L_{0}=\pi^{2}k_{B}/3e$ is shown in Fig.\
\ref{fig-lorentz-number}.  Deviations from the Wiedemann--Franz law
are found for $T \ge 0.1T_{K}$.  In the Kondo regime $L(T)$, like
$\kappa(T)$, is universal up to approximately $10T_{K}$. In the Kondo
regime it also exhibits a maximum at $T\approx T_{K}$ whereas in the
mixed valent there is only a shoulder at $k_BT\approx \Delta$, and
similarly in the empty orbital regime.  The inclusion of weak
non--resonant scattering has little effect on $\kappa(T)$.

\subsubsection*{Hall coefficient}

Fig.\ \ref{fig-hall-coefficient} shows the temperature dependence of
the Hall coefficient.  This quantity exhibits a low temperature peak
at $T\approx T_{K}$ in the Kondo regime but not in the mixed valent or
empty orbital regimes.  Universality for this quantity persists only
up to $T\approx T_{K}$.  Here we have only considered the ordinary
Hall effect and have shown that this exhibits an anomaly at $T\approx
T_{K}$. A more important contribution to the Hall effect arises from
skew scattering of conduction electrons. Up spin electrons are
scattered differently in a magnetic field than down spin electrons as
a result of the splitting of the Kondo resonance by the magnetic
field. This contribution should be much more important than that
coming from the Lorenz term considered here, but its treatment is
outside the scope of the present work.

\section*{Discussion and Conclusions}

In this paper we have given a detailed description of the
generalization of the numerical renormalization group to the
calculation of finite temperature Green's functions of the Anderson
model. The method was applied to obtain the single--particle spectral
densities and transport properties of the Anderson model in the strong
correlation limit $U/\pi\Delta >> 1$ for a range of temperatures and
parameter regimes, as well as the occupation number and specific heat.
We distinguish three parameter regimes in the limit of strong
correlation.  The Kondo regime, $\epsilon_{0}<<-\Delta,\;n_{0}\approx
1$, characterized by a low energy scale $k_{B}T_{0}=k_{B}T_{K}$, the
valence fluctuation regime, $-\Delta \le \epsilon_{0} \le 0,\; 0.3 \le
n_{0} \le 0.8$, characterized by a low energy scale
$k_{B}T_{0}=\Delta$, and the empty orbital regime, $\epsilon_{0}\ge
\Delta,\;n_{0}<0.3$, characterized by an energy scale
$k_{B}T_{0}\sim\tilde{\epsilon_0}\sim\epsilon_{0}$.  We also
distinguish three temperature ranges, $T<<T_{0}$ corresponding to the
Fermi liquid regime, $T>>T_{0}$ corresponding to the high temperature
perturbative regime and $T\approx T_{0}$ corresponding to the
cross--over region. A consistent picture of the thermodynamics,
spectral densities and transport coefficients emerges from the
numerical renormalization group calculations in these different
regimes.

The calculations of the single particle excitation spectrum show that
at low temperatures the spectral function in the Kondo regime is
characterized by a Kondo resonance of negligible spectral weight,
centered around $\epsilon_F$, and two atomic like resonances at
$\epsilon_{0}$ and $\epsilon_{0}+U$, which carry most of the spectral
weight. The one below the Fermi level at $\epsilon_{0}$ contains most
of the screening charge while the one above at $\epsilon_{0}+U$ is
empty.  This picture is well known and emerged very clearly in the
finite order perturbation calculations of Yamada \cite{yamada.75a}.
Using the numerical renormalization technique we have been able to
calculate it more accurately and obtain the correct exponential
dependence of the width $T_{K}$ on $\epsilon_{0}$ and $U$ and also
describe the high temperature regime where there are corrections that
depend logarithmically on $T$. \par The spectral function for valence
fluctuators is characterized by two peaks only: a partially filled one
at the renormalized position $\tilde{\epsilon_0}>\epsilon_{F}$ and an
empty one at $\tilde{\epsilon_0}+U$.  In the empty orbital regime the
low temperature spectral function in the strong coupling limit is just
an empty virtual bound state. The characteristic low energy scale for
the spectral function is set by $T_0$. These features of the spectral
density are fully consistent with the Fermi liquid groundstate as
revealed by the thermodyamics
\cite{krishnamurthy.80}. The Friedel sum rule which is a good test on the
accuracy of the method at both high and low energies was found to be
satisfied to within 3\% accuracy in all parameter regimes.  Increase
of temperature leads to a renormalization of the excitation spectrum
and for $T>T_0$ the spectral function changes in a qualitative way
\cite{horvatic.87}. At high temperatures the spectral function is
characterized by two atomic like peaks separated in energy by $U$,
consistent with the thermodynamic properties which exhibit local
moment behaviour. In the mixed valent and empty orbital regimes the
higher peak is strongly suppressed.

The transport coefficients for the Anderson model can be related to
the transport relaxation time by using linear response theory. This
can be further related to the single particle spectral function which
allows us to discuss the transport coefficients in terms of elementary
excitations in the resonant channel.  The characteristic low energy
scale of the spectral function, $T_0$, defines the temperature scale
for transport properties.

In the Kondo regime at low temperature, $T<<T_{0}=T_{K}$, we showed
that the transport coefficients agreed with the predictions of Fermi
liquid theory. For $T<<T_{K}$, the localized spins are screened and
the transport currents relax via excitations in the local Fermi liquid
state formed by the Kondo screening mechanism. The transport
coefficients were characterized by power law dependencies on $T/T_{K}$
(\ref{eq:scaling1}--\ref{eq:scaling4}) with universal coefficients for
appropriately scaled quantities. The numerical results for the $T^{2}$
term of the resistivity and the $T^{3}$ term of the thermal
conductivity in the symmetric case agreed remarkably well with the
analytic values deduced from Fermi liquid theory. With increasing
asymmetry the numerical results indicate that these coefficients
change but only slightly for $n_{0} > 0.8$. The thermopower shows
universal behaviour for $T\le T_{K}$ when rescaled by its zero
temperature slope \cite{zlatic.93}.  Universality was found to extend
up to at least $10T_{K}$ for the resistivity and thermal conductivity
but only up to $T_{K}$ for the Hall coefficient. The transport
properties in this regime are understood in terms of excitations in
the strongly renormalized Fermi liquid ground state. On increasing
temperature through the cross--over region $T\approx T_{K}$
logarithmic terms become important for the resistivity and these are
clearly seen for $T>T_{K}$. The thermopower changes sign, a result of
the disappearance for $T>T_{K}$ of the Kondo resonance in the
single--particle spectral density. There is also a minimum in the Hall
coefficient at $T\approx T_{K}$. In the high temperature perturbative
regime, $T>>T_{K}$, the spectral density consists of two peaks, the
occupancy $n_{0}$ approaches 1 and the transport properties are those
of electrons coupled weakly to local spins. The thermopower develops a
broad minimum at $k_BT\approx \Delta$ due to a shift of spectral
weight from the peak below the Fermi level to that above as charge
fluctuations become important.

In the mixed valent and empty orbital regimes the transport
coefficients at low temperature $T<<T_{0}$ again reflect the
relaxation of the transport currents via the excitations in the
renormalized local Fermi liquid ground state.  However, since $\Delta
>> k_BT_{K}$, these renormalizations of order $1/T_{0}$ are smaller
than for the Kondo case. The low temperature transport coefficients
again have power law dependencies on $T/T_0$, but the coefficients are
non--universal and depend strongly on the occupation of the localized
level $n_{0}$.

Finally, we investigated the effects of non--resonant scattering of
conduction electrons on the temperature dependence of the transport
properties. A general expression for the transport time in the
presence of such scattering was derived which shows that the resulting
interference effects between the resonant and non--resonant channels
are not negligible. Hence, to relate the transport coefficients of the
Anderson model to the experimental data on dilute alloys one should
take into account the non--resonant scattering. For example, the
overall shape of the thermopower curves calculated for the Anderson
model with two scattering channels resembles quite strikingly the
experimental data. The interference effects between resonant and
non--resonant scattering channels led to a strong dependence of the
sign of the thermopower on the non--resonant phase shift. This type of
effect could be important for systems with $n_{0}$ close to 1 or for
systems with a Kondo resonance approximately symmetric about the Fermi
level. Doping such systems with non--magnetic impurities could lead to
the above type of interference effects making the sign of the
thermopower sensitive to the local environment. Such effects have been
observed~\cite{bauer.91} In the symmetric Kondo regime the inclusion
of non--resonant scattering was important to obtain the non--vanishing
enhanced thermopowers for half--filled systems (the $n_{0}=1$).

To summarize, we have shown that the numerical renormalization group
method allows a unified description of the transport, thermodynamic
and spectral properties of the Anderson model in the strong
correlation limit. The different features below and above the
characteristic scale $T_0$ reflect differences in the structure of the
excitation spectrum at high and low temperatures. The NRG method
presented here could be used to provide further insights into the
behaviour of the Anderson model.  For instance, the results for the
spectral density could be used to deduce the spin and charge densities
induced in the conduction electrons at arbitrary distances from the
impurity \cite{szh}.  The interference effects for the transport
properties could also be studied for two Kondo (spin 1/2) impurities
or for the case of the Kondo effect in the $l=0$ channel and a virtual
bound state in the $l=1$ channel. This would be relevant to rare earth
systems in which there is an excited crystal electric field level
close to the Kondo ground state.  There are many other similarly
interesting physical situations which occur in particular materials.
They could be modelled and a broad range of physical properties
calculated via this one technique to compare with experiment.

\acknowledgments

We acknowledge the support of an SERC grant, and the Computational
Science Initiative for Computer equipment.  One of us (TAC) would like
to thank Prof. E.  M\"{u}ller--Hartmann for making possible a visit to
the Institute for Theoretical Physics at Cologne University where part
of this work was carried out.  Financial support from the {\em
Institute of Physics of the University of Zagreb}, (VZ,TAC), the
British Council (TAC) and the {\em Graduiertenkolleg
Festk\"{o}rperspektroskopie, University of Dortmund}, (TAC), is also
acknowledged.

\appendix
\section{Derivation of the transport time in the presence
of non--resonant scattering}

The transport time for scattering from a dilute concentration,
$n_{i}<<1$, of magnetic impurities described by the Hamiltonian
(\ref{eq:model}) can be obtained by examining the Kubo expression for
the conductivity $\sigma(T)$,

\begin{equation}
\sigma(T) = -lim_{\omega\rightarrow 0}{1\over \omega}\;\mbox{Im}\;
\Pi_{ret}(\omega),
\label{eq:conductivity}
\end{equation}
where the current--current correlation function $\Pi(i\omega)
=-\frac{1}{3V}\int_{0}^{\beta}e^{i\omega\tau}<T_{\tau}{\bf
j}(\tau){\bf j}(0)>$ and the current operator is that for
free--electrons ${\bf j}=-\frac{e}{m}\sum_{{\bf k}\sigma}{\bf
k}\;c_{{\bf k}\sigma}^{\dagger}c_{{\bf k}\sigma}$. The diagrammatic
representation of $\Pi(i\omega_{n})$ is shown in Fig.\
\ref{fig-diagram}.  In the absence of non--resonant scattering the
vertex corrections vanish identically \cite{bickers.87} in the dilute
limit leaving just the (dressed) bubble diagram $\Pi_{0}$. In terms of
the dressed conduction electron Green's function $G({{\bf k}},\omega)$
this is

\begin{equation}
\Pi_{0}(i\omega) = {{2e^{2}}\over {3m^{2}V}}
\sum_{{\bf k}}{\bf k}^{2}\frac{1}{\beta}\sum_{i\omega_{n}} G({\bf
k},i\omega_{n}+i\omega)G({\bf k},i\omega_{n}).
\end{equation}
Here we evaluate the conductivity including the vertex corrections
since these give rise to interference terms between the $l=0$ and
$l=1$ channels. The conductivity is given by \cite{mahan.81}

\begin{equation}
\sigma(T) = {1\over {2\pi}}\int_{-\infty}^{+\infty}
 \left( -\partial f \over
\partial\epsilon \right) \left[ P(\epsilon-i\delta,\epsilon+i\delta) -
Re[P(\epsilon+i\delta,\epsilon+i\delta)]\right] d\epsilon,
\end{equation}
where the singular part behaving like $1/n_{i}$ comes from the first
term with $P(\epsilon+i\delta,\epsilon-i\delta)$ \cite{mahan.81}. The
latter is given in terms of the conduction electron Green's function
$G({\bf k},\epsilon)$ and vector vertex function ${\bf \Gamma}$ by
\begin{equation}
P(\epsilon-i\delta,\epsilon+i\delta) = {{2e^{2}}\over {3m^{2}}}{1\over
V}
\sum_{k}G({\bf k},\epsilon+i\delta)G({\bf k},\epsilon-i\delta){\bf k}.{\bf
\Gamma}({\bf k},\epsilon-i\delta,
\epsilon+i\delta).\label{eq:p-formula}
\end{equation}
The vertex function ${\bf \Gamma}$ satisfies the following equation
\begin{equation}
{\bf \Gamma}({\bf k},\epsilon-i\delta,\epsilon+i\delta) = {\bf k} +
\int {{d^{3}{\bf k}'}\over (2\pi)^{3}}{\bf \Gamma}({\bf
k}',\epsilon-i\delta,\epsilon+i\delta) W_{{\bf
kk}'}(\epsilon-i\delta,\epsilon+i\delta)G({\bf k}',\epsilon-i\delta)
G(\epsilon+i\delta)\label{eq:vertex-equation}
\end{equation}
where $W$ is the irreducible vertex.

To solve this equation we introduce a scalar vertex function
$\gamma_{1}$ defined by ${\bf \Gamma}({\bf
k},\epsilon-i\delta,\epsilon+i\delta) ={\bf k}\;\gamma_{1}({\bf
k},\epsilon)$ and note that the Green's function product in
(\ref{eq:vertex-equation}) is just $A({\bf
k},\epsilon)/|\Sigma^{I}({\bf k},\epsilon)|$ where $A({\bf
k},\epsilon)$ is the spectral density and $\Sigma^{I}({\bf
k},\epsilon)$ the imaginary part of the self--energy of the conduction
electrons.  In the dilute limit, $n_{i}\rightarrow 0$, we also have
that \cite{mahan.81} $W_{{\bf
kk}'}(\epsilon-i\delta,\epsilon+i\delta)\rightarrow n_{i} T_{{\bf
kk}'}(\epsilon-i\delta)T_{{\bf k'k}}(\epsilon+i\delta)$, where $T$ is
the energy dependent T--matrix, $|\Sigma^{I}({\bf k}',\epsilon_{{\bf
k}'})|
\rightarrow n_{i}|Im T_{{\bf k'k'}}|$ and $A({\bf k}',\epsilon)\rightarrow 2\pi
\delta(\epsilon -\epsilon_{{\bf k}'})$. The resulting equation for the scalar
vertex function after some algebra becomes
\begin{equation}
\gamma_{1}({\bf k},\epsilon) = 1 + {m\over {4\pi}}\int d\epsilon_{{\bf
k}'}d\theta'
{\sqrt{2m\epsilon_{{\bf k}'}}}{{
\sin\theta'\cos\theta'\delta(\epsilon - \epsilon_{{\bf k}'})
T_{{\bf kk}'}(\epsilon-i\delta)T_{{\bf
k'k}}(\epsilon+i\delta)\gamma_{1}({\bf k}',\epsilon)} \over
{|Im\thinspace T_{{\bf
k'k'}}(\epsilon+i\delta)|}}\label{eq:scalar-vertex-equation}
\end{equation}
where $\theta'$ defined by $\cos(\theta')={\bf k.k'}/{\bf k}^{2}$ is
the angle between incoming and scattered k states and a free electron
density of states has been used to rewrite the integrals. The
T--matrix in the denominator of (\ref{eq:scalar-vertex-equation}) is
for forward scattering only so it is independent of $\theta'$ and can
therefore be taken outside the integral and evaluated at $k'=k$ (note
the $\delta$ function in (\ref{eq:scalar-vertex-equation})). The
scalar vertex function $\gamma_{1}({\bf k'},\epsilon)$ is also
independent of angle and can also be taken outside the integral
(evaluated at $k'=k$). Hence the closed expression,
\begin{equation}
\gamma_{1}(k,\epsilon) = 1 + {{mk}\over 4\pi}{I_{k}\over
{|Im\thinspace T_{kk}(\epsilon_{k}+i\delta)|}}\gamma_{1}(k,\epsilon),
\end{equation}
is obtained with solution,
\begin{equation}
\gamma_{1}(k,\epsilon_{k}) = {{|Im\thinspace
T_{kk}(\epsilon_{k}+i\delta)|}\over
{|Im\thinspace T_{kk}(\epsilon_{k}+i\delta)| - { {mk}\over{4\pi}
}I_{k} }},
\label{eq:scalar-vertex-solution}
\end{equation}
In the above expressions $I_{k}$ given by
\begin{equation}
I_{k} = \int_{0}^{\pi}d\theta' \sin\theta'\cos\theta' T_{{\bf
kk'}}(\epsilon_{{\bf k}}-i\delta)T_{{\bf k'k}}(\epsilon_{{\bf
k}}+i\delta)\label{eq:interference-integral}
\end{equation}
contains the interference effects.

Substituting the scalar vertex function $\gamma_{1}$ from
(\ref{eq:scalar-vertex-solution}) into the expression for
$P(\epsilon-i\delta,\;\epsilon+i\delta)$ (using the previosuly quoted
expression for the Green's function product) gives

\begin{equation}
P(\epsilon-i\delta,\epsilon+i\delta) = {{2e^{2}}\over {3m^{2}}}{1\over
V}
\sum_{{\bf k}}k^{2}A({\bf k},\epsilon)\gamma_{1}(k,\epsilon)/\Sigma^{I}({\bf
k},\epsilon)
\end{equation}
 and the latter into the conductivity formula (\ref{eq:conductivity})
gives for the transport time $\tau$ the expression

\begin{equation}
\tau(\epsilon_{k},T)^{-1} = 2n_{i} (|Im\thinspace T_{kk}(\epsilon_{k}+i\delta)|
-
{{mk}\over {4\pi}} I_{k}).\label{eq:transport-time-equation}
\end{equation}
The first part of this expression is just the usual result for s--wave
scattering only and the second part is the contribution coming from
interference effects between the resonant $l=0$ channel and the
non--resonant channels. To evaluate (\ref{eq:interference-integral})
and (\ref{eq:transport-time-equation}) we need the T--matrix. This can
be obtained for the Hamiltonian (\ref{eq:model}) by aplying the
equation of motion method to the conduction electron Green's function
$G_{{\bf kk}'\sigma}=<<c_{{\bf k}\sigma};c_{k'\sigma}^{\dagger}>>$.
This gives
\begin{equation}
G_{{\bf kk}'\sigma}(\omega+i\delta) = G_{{\bf
kk}'\sigma}^{0}(\omega+i\delta) + G_{{\bf
kk}'\sigma}^{0}(\omega+i\delta)T_{{\bf kk}'}(\omega+i\delta)G_{{\bf
kk}'\sigma}^{0}(\omega+i\delta),
\label{eq:eom-t-matrix}
\end{equation}
where
\begin{equation}
T_{{\bf kk}'}(\omega+i\delta) = V_{0{\bf
k}}G_{0}(\omega+i\delta)V_{0{\bf k}'} +
\sum_{l=1m}V_{lm{\bf k}}G_{l}(\omega+i\delta)V_{lm{\bf k}'}\label{eq:t-matrix},
\end{equation}
$G_{{\bf kk}'\sigma}^{0}$ is the unperturbed conduction electron
Green's function and $G_{l},l=0,1,\ldots$ are the local Green's
functions for the $l=0,1,\ldots$ channels respectively. The
hybridization matrix elements $V_{lm{\bf k}}$ have the angular
dependence $V_{lm{\bf k}} = <k|V|l,m> \sim V_{lk} Y_{lm}(\Omega_{{\bf
k}})$ so the $T$ matrix has the form:
\begin{eqnarray}
T_{{\bf kk}'}(\omega+i\delta) & = &
\sum_{l=0}(2l+1)P_{l}(\cos(\theta'))V_{lk}G_{l}(\omega+i\delta)V_{lk'}\\
& = & V_{0k}G_{0}(\omega+i\delta)V_{0k'} +
3\cos(\theta')V_{1k}G_{1}(\omega+i\delta)V_{1k'} + \dots\\
\end{eqnarray}
where $\sum_{m}Y_{lm}(\Omega_{k})Y_{lm}(\Omega_{k'}) \sim
(2l+1)P_{l}(\cos(\theta'))$ has been used and $\theta' = {\bf
k.k}'/{\bf k}^{2}$. The $T$ matrix has an explicit dependence on the
angle $\theta'$ which gives rise to interference effects between the
resonant and non--resonant channels when substituted into the
expression for the transport time (\ref{eq:transport-time-equation}).
These interference effects arise in the expression for $I_{k}$ as
cross--terms involving $G_{0}^{*}G_{1}+G_{0}G_{1}^{*}$.  On carrying
out the $\theta'$ integrations in (\ref{eq:interference-integral}) and
expressing the non--resonant Green's function $G_{1}(\omega)$ in terms
of a corresponding phase shift $\eta_{1}(\omega)$ we arrive at the
formula for the transport time of the multi--channel Anderson impurity
model:
\begin{equation}
\frac{1}{\tau(\omega,T)}
=
\frac{1}{\tau^{0}(\omega,T)}
\left[\cos{2\eta_1}
-
\frac{Re \;G_{0}(\omega,T)}{\mbox{Im}\; \;G_{0}(\omega,T)}\sin{2\eta_1}
\right]
+ \rho_n.\label{eq:tr-time}
\end{equation}
%
%
The resonant part of the transport time, $\tau_{\rm 0}(\omega,T)$ is
given by the usual $T$--matrix expression in the absence of the
interference effects,
\begin{equation}
\frac{1}{\tau_{0}(\omega,T)}
= \Delta\rho_{0}(\omega, T) = -\Delta\;\mbox{Im}\; G_{0}(\omega,T)/\pi
,
\end{equation}
where, $\Delta$ is the unrenormalized width of the $l=0$ resonance and
$G_0(\omega, T)$ is the $l=0$ Green's function, which describes the
many--body effects.  The analysis here was for the electrical
conductivity.  The same transport time (\ref{eq:tr-time}) is found by
repeating this analysis for the other current--current correlation
functions (In the above expressions for the inverse transport times, a
constant factor $2n_{i}/N(0)$, where $N(0)$ is the unperturbed
conduction electron density of states at the Fermi level, has been
omitted).

\section{The Numerical Renormalization Group Method}

The following continuum version for the resonant part of
(\ref{eq:model}) is useful for introducing the logarithmic
discretization approximation
\cite{krishnamurthy.80},
\begin{equation}
H = \epsilon_{0}c_{0\sigma}^{\dagger}c_{0\sigma} +
Un_{0{\uparrow}}n_{0{\downarrow}} +
V_{0}\int_{-1}^{+1}dk(c_{{k}\sigma}^{\dagger}c_{0\sigma} +
c_{0\sigma}^{\dagger}c_{{k}\sigma})+
\int_{-1}^{+1}kc_{k\sigma}^{\dagger}c_{k\sigma}dk\label{eq:continuum-model}.
\end{equation}
In obtaining this from the resonant part of (\ref{eq:model}), a
partial wave expansion for the conduction states around the impurity
has been used has been used and the Hamiltonian has been written in
the energy representation. The approximation has been made that
$\Delta(\omega)=\pi\sum_{k}|V_{0k}|^{2}\delta(\omega
-\epsilon_{k})=\Delta$ is independent of frequency so that only
s--wave partial wave states couple to the impurity (with strength
$V_{0}=\sqrt{\Delta/\pi N(\epsilon_{F})}$).  All energies are measured
relative to the half bandwidth $D=1$ and we use the convention that
repeated spin indices are summed over.  (full details of the above may
be found in \cite{krishnamurthy.80}). The continuum conduction band
$[-1,1]$ is now discretized as shown in Fig.\
\ref{fig-discrete-cond-band} and a new set of basis states for the
conduction electrons is introduced in each interval
$\pm[\Lambda^{-{(n+1)}} , \Lambda^{-n}]$ using the following
wavefunctions
\begin{equation}
\psi_{np}^{\pm}(k)  =  \left\{ \begin{array}{ll}
{\Lambda^{n/2} \over {D(1-1/\Lambda)^{1/2}}}e^{\pm i\omega_{n}pk} &
\mbox{for $\Lambda^{-n-1}< \pm k < \Lambda^{-n}$} \\
  0 & \mbox{otherwise}
\end{array} \right.
\end{equation}
Here p is a Fourier harmonic index and $\omega_{n}=2\pi
\Lambda^{n}/(1-\Lambda^{-1})$. The operators $c_{k\sigma}$ can then be expanded
in terms of new operators $a_{np\sigma},b_{np\sigma}$ labelled by the
interval $n$ and the harmonic index $p$
\begin{equation}
c_{{\bf
k}\sigma}=\sum_{np}[a_{np\sigma}\psi_{np}^{+}+b_{np\sigma}\psi_{np}^{-}].
\end{equation}
In terms of these operators, the Hamiltonian
(\ref{eq:continuum-model}) becomes~\cite{krishnamurthy.80},
\begin{eqnarray}
H & = & \epsilon_{0}c_{0\sigma}^{\dagger}c_{0\sigma} +
Un_{0{\uparrow}}n_{0{\downarrow}}\nonumber\\ & + &
V_{0}(1+\Lambda^{-1})^{1/2}\sum_{n=0}^{\infty}\Lambda^{-n/2}
((a_{n0\sigma}^{\dagger}+b_{n0\sigma}^{\dagger})c_{0\sigma} +
c_{0\sigma}^{\dagger}(a_{n0\sigma}+b_{n0\sigma}))\nonumber\\ & + &
{1\over 2}(1+\Lambda^{-1})
\sum_{np}\Lambda^{-n}(a_{np\sigma}^{\dagger}a_{np\sigma} -
b_{np\sigma}^{\dagger}b_{np\sigma})\nonumber\\
& + &{(1-\Lambda^{-1})\over {2\pi i}}\sum_{n}\sum_{p\ne
p'}\Lambda^{-n} (a_{np\sigma}^{\dagger}a_{np'\sigma} -
b_{np\sigma}^{\dagger}b_{np'\sigma}) e^{{{2\pi i (p-p')} \over
(1-\Lambda^{-1})}}.\label{eq:log-form}
\end{eqnarray}
The coupling of the local state $c_{0\sigma}$ to just the $p=0$
harmonic is a consequence of the assumption that the hybridization
matrix elements are independent of energy. Hence the conduction
electron orbitals $a_{np},b_{np}$ for $p\ne 0$ only couple to the
impurity state $c_{0\sigma}$ indirectly via their coupling to the
$a_{n0},b_{n0}$ in the last term of (\ref{eq:log-form}).  This
coupling is weak, being proportional to $(1-\Lambda^{-1})$, and
vanishes in the continuum limit $\Lambda\longrightarrow 1$, so these
states may be expected to contribute little to the impurity properties
compared to the $p=0$ states.  This is indeed the case as can be seen
in Fig.\ \ref{fig-psi-orbitals}, where the probability density in real
space of the wave--packet states $\psi_{np}(k)$ is shown. It can be
seen that the $p>0$ orbitals are localized at increasing distances
from the impurity, whereas the $p=0$ orbitals overlap strongly with
the impurity. This together with their weak coupling to the impurity
ensures that they can be neglected in calculating local impurity
properties. This {\em logarithmic discretization approximation} has
been shown to give rapidly convergent results (as
$\Lambda\longrightarrow 1$) for thermodynamic averages.  The
deviations away from the continuum limit, $\Lambda=1$, being
proportional to $e^{-\frac{\pi^2}{\ln(\Lambda)}}$
\cite{wilson.75,krishnamurthy.80}.  The approximation amounts to
approximating the continuum conduction band Hamiltonian $H_{c}$, by a
discrete one:
\begin{equation}
H_{c} \approx {1\over 2}(1+\Lambda^{-1})
\sum_{n}\Lambda^{-n}(a_{n0\sigma}^{\dagger}a_{n0\sigma} -
b_{n0\sigma}^{\dagger}b_{n0\sigma}).\label{eq:discrete-band}
\end{equation}
Consequently a discrete approximation to the Anderson model is
obtained
\begin{eqnarray}
H & = & \epsilon_{0}c_{0\sigma}^{\dagger}c_{0\sigma} +
Un_{0{\uparrow}}n_{0{\downarrow}}\nonumber\\ & + &
V_{0}(1+\Lambda^{-1})^{1/2}\sum_{n=0}^{\infty}\Lambda^{-n/2}
((a_{n\sigma}^{\dagger}+b_{n\sigma}^{\dagger})c_{0\sigma} +
c_{0\sigma}^{\dagger}(a_{n\sigma}+b_{n\sigma}))\nonumber\\ & + &
{1\over 2}(1+\Lambda^{-1})
\sum_{n}\Lambda^{-n}(a_{n\sigma}^{\dagger}a_{n\sigma} -
b_{n\sigma}^{\dagger}b_{n\sigma}),\label{eq:discrete-model}
\end{eqnarray}
where the p index has now been dropped since only $p=0$ states are
included.  This discrete Hamiltonian is more convenient for a
numerical treatment.

The next step is to convert the discrete Hamiltonian
(\ref{eq:discrete-model}) into a linear chain form suitable for an
iterative diagonalization starting from a small chain and
diagonalizing sucessively longer chains by adding a site or energy
shell at each stage.  If we define a new local conduction electron
orbital by $|0\sigma>=f_{0\sigma}^{\dagger}|\mbox{vac}>$ where
\begin{equation}
f_{0\sigma}^{\dagger}=[{1\over
2}(1+\Lambda^{-1})]^{1/2}\sum_{n=0}^{\infty}\Lambda^{-n/2}
((a_{n\sigma}^{\dagger}+b_{n\sigma}^{\dagger}),
\end{equation}
then the Lanczos algorithm can be used with starting vector
$|0\sigma>$ to tri--diagonalize $H_{c}$
\begin{equation}
H_{c} = \sum_{n=0}^{\infty}\epsilon_{n}
f_{n\sigma}^{\dagger}f_{n\sigma}+
\sum_{n=0}^{\infty}\lambda_{n}(f_{n+1\sigma}^{\dagger}f_{n\sigma}+f_{n\sigma}^{\dagger}f_{n+1\sigma}),\label{eq:band-chain}
\end{equation}
where $\epsilon_n, \lambda_n, n=0,1,\ldots$ are the site and hopping
energies respectively. For the half--filled Fermi sea $|\mbox{vac}>$
the $\epsilon_{n}$ are all zero and Wilson has obtained the expression
$\lambda_{n}= {1\over 2}(1+\Lambda^{-1})\Lambda{^{-n/2}}\xi_{n}$,
where $\xi_{n} =
\xi_{n}(\Lambda)\rightarrow 1$ for $n>>1$.  This converts
(\ref{eq:discrete-model}) into a semi--infinite chain with the
impurity at the origin coupled to the local conduction state
$|0\sigma>$ via the hybridization $V_{0}$ (re--scaled to keep
$|0\sigma>$ normalized)
\begin{eqnarray}
H & = & \epsilon_{0}c_{0\sigma}^{\dagger}c_{0\sigma} +
Un_{0\uparrow}n_{0\downarrow} + V_{0}(f_{0\sigma}^{\dagger}c_{0\sigma}
+ H.c.)\nonumber\\ & + & {1 \over 2}(1+\Lambda^{-1})
\sum_{n=0}^{\infty}\lambda_{n}(f_{n+1\sigma}^{\dagger}f_{n\sigma}+f_{n\sigma}^{\dagger}f_{n+1\sigma})\label{eq:discrete-chain-model}\\
& = & lim_{N\rightarrow \infty}{1\over 2}(1+\Lambda^{-1})
\Lambda^{-(N-1)/2}H_{N}\nonumber\\
H_{N} & = & \Lambda^{(N-1)/2}[ H_{0} + H_{hyb} +
\sum_{n=0}^{N-1}\Lambda^{-n/2}\xi_{n}(f_{n+1\sigma}^{\dagger}
f_{n\sigma}+f_{n\sigma}^{\dagger}f_{n+1\sigma})].
\label{eq:rescaled-form}
\end{eqnarray}
In (~\ref{eq:rescaled-form}) we have defined a sequence of rescaled
Hamiltonians $H_{N}$ whose smallest term is always of order 1 (the
bare parameters $\epsilon_{0},U,V_{0}$ are also rescaled by the factor
${1\over 2}(1+\Lambda^{-1})$). The procedure is now to diagonalize
this sequence of Hamiltonians iteratively and extract the spectrum and
eigenstates on successively lower energy scales $\omega_{N} \sim
\Lambda^{-(N-1)/2}$.  The procedure starts by diagonalizing the
impurity part, $H_{0}=\epsilon_{0}c_{0\sigma}^{\dagger}c_{0\sigma} +
Un_{0\uparrow}n_{0\downarrow}$ of (\ref{eq:discrete-chain-model}) and
then adding the coupling to the local conduction electron orbital
$V_{0}(f_{0\sigma}^{\dagger}c_{0\sigma} + H.c.)$. Successive energy
shells $\lambda_{n}(f_{n+1\sigma}^{\dagger}f_{n\sigma}
+f_{n\sigma}^{\dagger}f_{n+1\sigma}),n=0,1,\ldots $ are then added and
the resulting Hamiltonian diagonalized to give the many--body
eigenvalues $E_{p}^{N}$ and eigenvectors $|p>_{N}$ of the
corresponding shell or cluster.  The total number of electrons,
$N_{e}$, total spin, $S$, and z--component of total spin, $S_{z}$, are
conserved quantities and can be used to label the eigenstates together
with an index $r=1,R_{N_{e}S}^{N}$ where $R_{N_{e}S}^{N}$ is the
dimension of the subspace $(N_{e}S)$. If $H_{N}$ has been
diagonalized, $H_{N}=\sum_{N_{e}SS_{z}r}E_{N_{e}S}^{N}X_{N_{e}SS_{z}r,
N_{e}SS_{z}r}^{N}$, then from (\ref{eq:rescaled-form}) the matrix for
$H_{N+1}$ can be obtained from
\begin{equation}
H_{N+1} = \Lambda^{1/2}H_{N} +
\xi_{N}(f_{N+1\sigma}^{\dagger}f_{N\sigma}
+f_{N\sigma}^{\dagger}f_{N+1\sigma}).\label{eq:recursion-relation}
\end{equation}
The basis used for $H_{N+1}$ is the product basis $|p,i>$ of
eigenstates, $|p>=|N_{e}SS_{z}r>$, of $H_{N}$, and states $|i>$ from
site $N+1$ (i.e.  $|i>=|0>, |\uparrow>, |\downarrow>,
|\uparrow\downarrow>$).  In this basis the matrix elements of
$H_{N+1}$ are
\begin{eqnarray}
<p,i|H_{N+1}|q,j> & = & \Lambda^{1/2}\delta_{p,q}\delta_{i,j}E_{p}^{N}
+
\xi_{N}(<p,i|f_{N+1\sigma}^{\dagger}f_{N\sigma}|q,j> +
<p,i|f_{N\sigma}^{\dagger}f_{N+1\sigma}|q,j>)\nonumber\\ & = &
\Lambda^{1/2}\delta_{p,q}\delta_{i,j}E_{p}^{N}\nonumber\\ & + &
\xi_{N}(-1)^{N_{e,q}}(<i|f_{N+1\sigma}^{\dagger}|j><p|f_{N\sigma}|q> +
<p|f_{N\sigma}^{\dagger}|q><i|f_{N+1\sigma}|j>),
\label{eq:ham-matrix-elements}
\end{eqnarray}
where $f_{N+1\sigma}$ have been commuted past the eigenstates $|p>,
|q>$ with a sign change $(-1)^{N_{e,q}}$ depending on the number of
electrons, $N_{e,q}=N_{e,p}$, in these states. Since $H_{N}$ is
already diagonalized the matrix elements $<p|f_{N\sigma}|q>$ can be
calculated as follows.  The unitary transformation between the
eigenstates of $H_{N}$ and the product basis is known and is given by
\begin{equation}
|p>_{N} = \sum_{r,i}U_{N}(p,ri)|r>_{N-1}|i>,\label{eq:unitary-trans}
\end{equation}
where $U_{N}(p,ri)$ is the matrix of eigenvectors of $H_{N}$,
$|r>_{N-1}$ denotes an eigenvector of the previous cluster ($N-1$) and
$|i>$ is one of the four states given above, but for site $N$ (not
$N+1$).  Hence the required matrix elements are given by
\begin{eqnarray}
<p|f_{N\sigma}|q>_{N} & =
&\sum_{r,i}\sum_{r',i'}U_{N}^{*}(p,ri)U_{N}(p,r'i')
<i|<r|f_{N\sigma}|r'>|i'>\nonumber\\ & = &
\sum_{r,i}\sum_{r',i'}U_{N}^{*}(p,ri)U_{N}(q,r'i')
\delta_{r,r'}(-1)^{N_{e,r}}<i|f_{N\sigma}|i'>,
\label{eq:f-matrix-elements}
\end{eqnarray}
which involves only known expressions.  The matrix elements
$<i|f_{N+1\sigma}|j>$ in (\ref{eq:ham-matrix-elements}) are also known
so the matrix of $H_{N+1}$ can be set up in terms of the eigenvalues
and matrix elements in (\ref{eq:f-matrix-elements}). $H_{N+1}$ can
then be diagonalized and the procedure repeated to obtain the spectrum
on successively lower energy scales, $\omega_{N+2}, \omega_{N+3},
\ldots$, where $\omega_{N} \approx D\Lambda^{-{{N-1}\over {2}}}$ is
the smallest scale in $H_{N}$. In practice since the number the number
of many--body states in $H_{N}$ grows like $4^{N}$ it is not possible
to retain all states after about $N=7$ (even after symmetry has been
used to reduce the size of the matrices). For $N>7$ only the lowest
1000 or so states of $H_{N}$ are retained. The truncation of the
spectrum in $H_{N}$ restricts the reliable range of excitations
$\omega$ to $\omega_{N} \le
\omega
\le K \omega_{N}$ where K is a constant dependent on $\Lambda$
($K\approx 10$ for $\Lambda=3$). The lower excitations are calculated
more reliably in successive iterations, whilst information on the
higher excitations is contained in previous iterations.  The
eigenvalues are used to calculate the partition function
$Z_{N}(\beta_{N}=1/k_{B}T_{N})$ and free energy
$F(T_{N})=-k_{B}\ln(Z_{N}(T_{N}))$ for the total system at a
decreasing sequence of temperatures $T_{N} = \omega_{N}/k_{B} \sim
D\Lambda^{-(N-1)/2}, N=0,1,\ldots$. The thermodynamic properties are
then extracted by first subtracting out the conduction electron
contribution to the free energy, $F_{c}(T_{N})$
\cite{wilson.75,krishnamurthy.80}. The specific heat is then given
by the second derivative of the impurity free energy,
$F_{imp}(T)=F(T)-F_{c}(T)$,
\begin{equation}
C(T) = -T\frac{\partial^{2}F_{imp}}{\partial T^{2}}
\end{equation}

\newpage
\begin{figure}
\caption{
The local level ocuupancy $n_{0}(T)$ over the whole temperature range
in various regimes.  }
\label{fig-valence}
\end{figure}
\begin{figure}
\caption{
The specific heat in units of $k_{B}$ over the whole temperature range
and in the Kondo, mixed valent and empty orbital regimes.  }
\label{fig-specific-heat-all}
\end{figure}
\begin{figure}
\caption{
The specific heat exhibits Fermi liquid behaviour, $C(T)=\gamma T$, at
low temperature, $T<<T_{K}$, in all parameter regimes. Note the very
much enhanced values of $\gamma$ in the Kondo cases compared to the
mixed valent and empty orbital case.  }
\label{fig-specific-heat-gamma}
\end{figure}
\begin{figure}
\caption{
The specific heat in the Kondo regime, showing the universal behaviour
for $T\le T_{K}$. However, when the local level position is well
separated from the Fermi level, universality extends up to almost
$10T_{K}$ over almost the entire range of the low energy peak. The two
disticnt energy scales, $T_{K}$, governing the low energy peak and
$\Delta>>T_{K}$ governing the high energy charge fluctuation peak are
clearly evident.  }
\label{fig-specific-heat-universal}
\end{figure}
\begin{figure}
\caption{
The $T=0$ spectral density in the Kondo regime for several positions
of the local level: the lower satellite peak is at the local level
position $\epsilon_{0}$. This takes values $\epsilon_{0}=-U/2$
(symmetric case), $\epsilon_{0}/\Delta =-4, -3, -2$ (asymmetric
cases).  For $\epsilon_{0}/\Delta = -2$ the lower satellite peak forms
a shoulder and has partly merged with the Kondo resonance. The high
energy satellite peak is separated from the lower one by the Coulomb
energy U. The inset shows the Kondo resonance in more detail. The
Kondo resonance becomes broader as $-\epsilon_{0}$ decreases.  }
\label{fig-spec-zero-kondo}
\end{figure}
\begin{figure}
\caption{
The finite temperature spectral density for the symmetric case. The
inset is for $\rho(\omega,T)$ plotted versus $T/T_{K}$ and shows in
more detail the region around the Fermi level. The satellite peaks
acquire some temperature dependence for $T\approx \Delta$.  }
\label{fig-spec-finite-symm-kondo}
\end{figure}
\begin{figure}
\caption{
The spectral densities for the asymmetric case in the Kondo regime.
At high temperatures $T\ge \Delta$ there is a shift of spectral weight
from the lower to the upper satellite peak. The Kondo resonance, shown
more clearly in the inset where it is plotted versus $T/T_{K}$,
disappears for $T>>T_{K}$.  }
\label{fig-spec-finite-asym-kondo}
\end{figure}
\begin{figure}
\caption{
The $T=0$ spectral densities in the mixed valent
($\epsilon_{0}/\Delta=-1,0$) and empty orbital
($\epsilon_{0}/\Delta=+1$) regimes. The resonant level of width of
order $\Delta$ is renormalized by the interactions to lie above the
Fermi level. There is no Kondo resonance in these cases and the upper
satellite peak at approximately $\epsilon_{0}+U$ has very little
weight.  }
\label{fig-spec-zero-mixed+empty}
\end{figure}
\begin{figure}
\caption{
The temperature dependence of the spectral densities in the mixed
valent regime.  The renormalized resonant level of width $\Delta$
lying at $\tilde{\epsilon_{0}}>0$ acquires a strong temperature
dependence on a scale of $T\approx \Delta$.  }
\label{fig-spec-finite-mixed-valent}
\end{figure}
\begin{figure}
\caption{
The temperature dependence of the spectral densities in the empty
orbital regime. The relevant scale is the renormalized level
$\tilde{\epsilon_{0}}>0$ which for the present choice of parameters is
approximately $\Delta$.  On a scale of
$\tilde{\epsilon_{0}}\approx\Delta$ this resonance has a strong
dependence on temperature.  }
\label{fig-spec-finite-empty-orbital}
\end{figure}
\begin{figure}
\caption{
The electrical resistivity in various parameter regimes and over the
whole temperature range. The curves are plotted versus the reduced
temperature $T/T_{0}$, where $T_{0}=\Delta$ in the mixed valent and
empty orbital regimes and $T_{0}=T_{K}$ in the Kondo regime. The zero
temperature value, $\rho(T=0)$, is fixed by the Friedel sum rule to be
$\sin^{2}(\pi n_{0}(T=0)/2)/\pi \Delta$.  }
\label{fig-resistivity-all}
\end{figure}
\begin{figure}
\caption{
The resistivity in the symmetric case with a fit of the Hamann
expression with $T_{KH}=T_{K}/1.2$. The Hamann result fits the high
temperature range, and is very close to the NRG data for $T_{K}\le T
\le 10T_{K}$. It fails at low and very high temperatures.  }
\label{fig-resistivity-hamann}
\end{figure}
\begin{figure}
\caption{
The scaled resistivity in the Kondo regime showing the universal
behaviour at low temperature up to approximately $5T_{K}$. The inset
for $1-(\rho(T)/\rho(0))$ versus $(T/T_{K})^{2}$ shows the expected
Fermi liquid behaviour for the resistivity at low temperature
$T<0.1T_{K}$. The coefficient, $c$, of the $T^{2}$ term in the
resistivity is found to lie within 8 \% of the exact result in all
cases.  }
\label{fig-resistivity-universal+fliq}
\end{figure}
\begin{figure}
\caption{
The influence of non--resonant scattering of conduction electrons on
the resistivity in the symmetric Kondo regime.}
\label{fig-resistivity-nonres-symm}
\end{figure}
\begin{figure}
\caption{
$S(T)/(T/T_{K})$ showing the linear in $T$ Fermi liquid behaviour of
the thermopower for $T<<T_{K}$. The different curves are for local
level positions $\epsilon_{0}=-4\Delta (\circ )$,
$\epsilon_{0}=-3\Delta (\Box )$, $\epsilon_{0}=-2\Delta (\diamond )$,
$\epsilon_{0}=-1\Delta (\bigtriangleup )$, $\epsilon_{0}= 0
(\bigtriangledown )$, $\epsilon_{0}=+1\Delta (+)$. The inset shows the
first three cases in more detail.  }
\label{fig-S-low}
\end{figure}
\begin{figure}
\caption{
The thermopower $S(T)$ in the Kondo regime. The low energy maximum is
at a temperature $T$ in the range $T_{K}/3\le T\le T_{K}$ whilst the
broad minimum at higher temperature is at $T\approx \Delta$.  }
\label{fig-thermo-kondo}
\end{figure}
\begin{figure}
\caption{
The thermopower $S(T)$ in the mixed valent and empty orbital regimes.
The low energy maximum in the mixed valent case is in the range
$\Delta/3\le T\le
\Delta$ and the minimum at higher temperature is at $T\approx \Delta$. In the
empty orbital case, there is only a shoulder at $T\approx \Delta$ }
\label{fig-thermo-mixed}
\end{figure}
\begin{figure}
\caption{
The thermopower $S(T)$ in the symmetric case with inclusion of
non--resonant scattering.  The low energy minimum is at $T\approx
T_{K}$ and the extremum at higher temperature is at $T\approx
50T_{K}\approx 0.9\Delta$.  }
\label{fig-thermo-symm}
\end{figure}
\begin{figure}
\caption{
The effect of including non--resonant scattering on the thermopower in
the Kondo regime $\epsilon_{0}=-2\Delta$. For a large enough
non--resonant phase shift $\eta_{1}>0.042\pi$ the broad minimum at
$T\approx \Delta$ changes sign. The low energy maximum is at $T\approx
T_{K}/3$.  }
\label{fig-thermo-kondo+nr}
\end{figure}
\begin{figure}
\caption{
The effect of including non--resonant scattering on the thermopower
$S(T)$ in the mixed valent regime $\epsilon_{0}=0$.  There is only a
broad maximum at $T\approx \Delta/3$ which disappears when the
non--resonant phase--shift $\eta_{1}$ is increased.  }
\label{fig-thermo-mixed+nr}
\end{figure}
\begin{figure}
\caption{
The thermal conductivity $\frac{(\kappa(T)/T)}{(\kappa(T)/T)_{0}}$
plotted versus $\gamma\thinspace T$. The solid lines with symbols are
for the Kondo regime. The mixed valent cases are $\epsilon_{0}
=-\Delta$ (long dashed) and $\epsilon_{0} = 0$ (dashed), and
$\epsilon_{0} =-\Delta$ (dotted) is the empty orbital case.  (The
single point for the symmetric case which falls off the universal
curve is due insufficient accuracy of the integrations at that
particular temperature)}
\label{fig-thermal-cond}
\end{figure}
\begin{figure}
\caption{
The linear coefficient of the thermal conductivity $\alpha =
(\kappa(T)/T)_{0}$ versus $\frac{1}{\sin^{2}(\pi n_{0}/2)}$ (see Table
I for the values).  }
\label{fig-thermal-cond-alpha}
\end{figure}
\begin{figure}
\caption{
The $T^{2}$ coefficient of $\kappa(T)/\alpha\thinspace T$ in the Kondo
regime (see Table I for the values).  }
\label{fig-thermal-cond-beta}
\end{figure}
\begin{figure}
\caption{
The Lorenz number ratio $L(T)/L_{0}$ where $L_{0}=\pi^{2}k_{B}/3e$
versus $\gamma\thinspace T$.  }
\label{fig-lorentz-number}
\end{figure}
\begin{figure}
\caption{
The Hall coefficient $R(T)/R_{inf}$ versus $\gamma\;T$ where
$R_{inf}=-1/n|e|c$.  }
\label{fig-hall-coefficient}
\end{figure}
\begin{figure}
\caption{
The diagrammatic representation of the current--current correlation
function $\Pi(i\omega)$. The solid lines represent full conduction
electron Green's functions and $\Gamma$ is the two--particel
scattering vertex.  }
\label{fig-diagram}
\end{figure}
\begin{figure}
\caption{
The conduction band $[-1,+1]$ logarithmically discretized with
discretization parameter $\Lambda>1$ }
\label{fig-discrete-cond-band}
\end{figure}
\begin{figure}
\caption{
The normalized conduction electron orbitals $|k_{F}r\psi_{np}(r)|^{2}$
as a function of distance $k_{F}r$ from the impurity for $n=5$ and
$p=0,1,\dots,5$.  }
\label{fig-psi-orbitals}
\end{figure}
\newpage
\squeezetable
\mediumtext
\begin{table}
\caption
{The first column shows the local level positions used in the
calculations, and the other columns show various data extracted.
$\rho_{0}^{Friedel}={ {\sin^{2}(\pi n_{0}/2)} \over {\pi\Delta}}$,
where $n_{0}$ is the local level occupancy at $T=0$ and is extracted
from the partition function. The $T^{2}$ coefficient of the
resistivity in the Kondo regime is defined by $c = lim_{T\rightarrow
0} \left( {1-\rho(T)/\rho(0)} \over {(T/T_{K})^{2}} \right)$, which
according to Fermi liquid theory should be $6.088$ in the symmetric
case with small corrections in the asymmetric Kondo regime.  The
linear coefficient of the thermal conductivity is $\alpha=\left(
\kappa(T)\over T
\right)_{T\rightarrow 0}$ and $\beta$ is the $T^{2}$ coefficient of
$\kappa(T)/\alpha T$ which should be $15.83$ in the symmetric case.  }
\label{table1}
\begin{tabular}{ccccccccccc}
$\epsilon_{0}$ & $n_{0}$ & $\gamma/k_{B}^{2}$ & $T_{K}/\Delta$
&$\rho_{0}^{Friedel}$ &$\rho_{0}^{NRG}(\epsilon_{F},0)$ & \% error
&$c$ & \% error & $\alpha$ & $\beta$\\
\tableline
$\epsilon_{0}=-{U \over 2}$& 1.00 & 8055.5 & 0.0180 & 31.83 & 32.61 &
-2.4\% & 5.7 & -6.4\% & 0.102 & $ 11.6 $
\\					 $\epsilon_{0}=-4\Delta$ &
0.93 & 4224.0 & 0.0346 & 31.45 & 32.13 & -2.1\% & 5.8 & -4.7\% & 0.101
& $ 13.5 $\\
$\epsilon_{0}=-3\Delta$ & 0.88 & 2069.0 & 0.0690 & 31.26 & 31.33 &
-0.2\% & 6.4 & +5.1\% & 0.106 & $ 12.6
$\\					 $\epsilon_{0}=-2\Delta$ &
0.78 & 832.0 & 0.1790 & 28.34 & 28.93 & -2.0\% & 6.6 & +8.4\% & 0.113
& $ 13.2 $\\
$\epsilon_{0}=-\Delta$ & 0.63 & 281.0 & 0.5900 & 22.14 & 22.32 &
-0.8\% & & & 0.147 & \\
$\epsilon_{0}=0$ & 0.45 & 100.7 & 2.5100 & 13.18 & 13.36 & -1.4\% & &
& 0.246 &
\\					 $\epsilon_{0}=+\Delta$ & 0.31 & 42.2 &
13.660 & 6.85 & 6.79 & +0.9\% & & & 0.496 & \\
\end{tabular}
\end{table}

\end{document}